\documentclass[a4paper,11pt]{article}
\usepackage{jinstpub} 
\usepackage{lineno}

\usepackage{booktabs}
\usepackage{tabularx}

\newcommand{\mpi}{M_\pi}
\newcommand{\mpii}{M_{\pi^0}}
\newcommand{\<}{\langle}
\renewcommand{\>}{\rangle}
\newcommand{\nn}{\nonumber\\}
\newcommand{\MeV}{\,\text{MeV}}
\newcommand{\GeV}{\,\text{GeV}}
\renewcommand{\Re}{\mathrm{Re}}
\renewcommand{\Im}{\mathrm{Im}}

\preprint{
\mbox{}\hfill{} PSI-PR-23-28 \\
\mbox{}\hfill{} ZH-TH 42/23
}

\title{\boldmath Puzzles in the hadronic contributions to the muon anomalous magnetic moment}

\author[a,1
]{Gilberto Colangelo,\note{Speaker at ``21st Conference on Flavor Physics and CP Violation,'' 2023, Lyon, France.}%
}

\author[a,2
]{Martin Hoferichter,\note{Speaker at ``Les Rencontres de Physique de la Vall\'ee d’Aoste,'' 2023, La Thuile, Italy.}%
}
\author[b,c,3
]{Peter Stoffer,\note{Speaker at ``New Frontiers in Lepton Flavor,'' 2023, Pisa, Italy.}%
}

\affiliation[a]{Albert Einstein Center for Fundamental Physics, Institute for Theoretical Physics, University of Bern, Sidlerstrasse 5, 3012 Bern, Switzerland}
\affiliation[b]{Physik-Institut, Universit\"at Z\"urich, Winterthurerstrasse 190, 8057 Z\"urich, Switzerland}
\affiliation[c]{Paul Scherrer Institut, 5232 Villigen PSI, Switzerland}

\emailAdd{gilberto@itp.unibe.ch}
\emailAdd{hoferichter@itp.unibe.ch}
\emailAdd{stoffer@physik.uzh.ch}

\abstract{We summarize recent developments in the Standard-Model evaluation of the anomalous magnetic moment of the muon $a_\mu$, both in the hadronic-light-by-light and hadronic-vacuum-polarization contributions. The current situation for the latter is puzzling as we are confronted with multiple discrepancies that are not yet understood. We present updated fits of a dispersive representation of the pion vector form factor to the new CMD-3 data set and quantify the tensions with the other high-statistics $e^+e^-\to\pi^+\pi^-$ experiments in the contribution to $a_\mu$ in the energy range up to $1\GeV$, as well as in the corresponding contribution to the intermediate Euclidean window.}

\begin{document}
\maketitle
\flushbottom

\section{Introduction}
\label{sec:Intro}

The anomalous magnetic moment of the muon $a_\mu = (g-2)_\mu/2$ has received a lot of attention over the past years, as the current experimental value~\cite{Muong-2:2021ojo,Muong-2:2021ovs,Muong-2:2021xzz,Muong-2:2021vma,Muong-2:2006rrc}
\begin{align}
	\label{eq:amuExp}
	a_\mu^\text{exp}=116\,592\,061(41)\times 10^{-11}
\end{align}
already has an impressive precision of $0.35\,\text{ppm}$ and further improvements from the Fermilab experiment are expected in the near future. In order to fully exploit this progress on the experimental side, the theoretical prediction within the Standard Model (SM) needs to achieve a similar level of precision. The experimental value~\eqref{eq:amuExp} is in $4.2\sigma$ tension with the SM prediction~\cite{Aoyama:2012wk,Aoyama:2019ryr,Czarnecki:2002nt,Gnendiger:2013pva,Davier:2017zfy,Keshavarzi:2018mgv,Colangelo:2018mtw,Hoferichter:2019gzf,Davier:2019can,Keshavarzi:2019abf,Hoid:2020xjs,Kurz:2014wya,Melnikov:2003xd,Colangelo:2014dfa,Colangelo:2014pva,Colangelo:2015ama,Masjuan:2017tvw,Colangelo:2017qdm,Colangelo:2017fiz,Hoferichter:2018dmo,Hoferichter:2018kwz,Gerardin:2019vio,Bijnens:2019ghy,Colangelo:2019lpu,Colangelo:2019uex,Blum:2019ugy,Colangelo:2014qya}
\begin{align}
	\label{eq:amuSM}
	a_\mu^\text{WP} = 116\,591\,810(43)\times 10^{-11} \, ,
\end{align}
as published in the 2020 White Paper (WP)~\cite{Aoyama:2020ynm}. The theoretical uncertainty is completely dominated by hadronic effects, in particular by hadronic vacuum polarization (HVP), which in Eq.~\eqref{eq:amuSM} is determined via dispersion relations and experimental input on the photon-inclusive $e^+e^-\to\text{hadrons}$ cross sections according to~\cite{Bouchiat:1961lbg,Brodsky:1967sr} 
\begin{align}
	\label{eq:amuHVP}
	a_\mu^\text{HVP, LO} &= \bigg(\frac{\alpha m_\mu}{3\pi}\bigg)^2\int_{s_\text{thr}}^\infty ds \, \frac{\hat K(s)}{s^2}R_\text{had}(s) \,, \quad R_\text{had}(s) = \frac{3s}{4\pi\alpha^2}\sigma(e^+e^-\to\text{hadrons}(+\gamma)) \, . 
\end{align} 
The QED kernel function $\hat K(s)$ is known analytically and the integration starts at the $\pi^0\gamma$ threshold $s_\text{thr}=\mpii^2$. Achieving sub-percent accuracy in the HVP evaluation requires sufficient control over radiative corrections in the hadronic $R$-ratio $R_\text{had}$. The WP result~\eqref{eq:amuSM} based on compilations of $e^+e^-$ measurements has been challenged
by the first lattice-QCD result achieving sub-percent precision~\cite{Borsanyi:2020mff}. Moreover, the new $e^+e^-$ data set on the two-pion channel by CMD-3~\cite{CMD-3:2023alj} differs significantly from the input used in the WP.

Here, we summarize the current status of the SM prediction for $a_\mu$, which now needs to address a multitude of discrepancies and tensions (see Ref.~\cite{MuonInitiative}). We focus on the hadronic contributions, describing recent progress on the sub-leading hadronic light-by-light (HLbL) scattering in Sect.~\ref{sec:HLbL}, before discussing several aspects of HVP in Sect.~\ref{sec:HVP}.

\section{Hadronic light-by-light scattering}
\label{sec:HLbL}

The HLbL contribution to $a_\mu$ is determined by the hadronic four-point correlator of electromagnetic currents
\begin{align}
	\Pi^{\mu\nu\lambda\sigma}&(q_1,q_2,q_3) \nn
		&= -i \int d^4x \, d^4y \, d^4z \, e^{-i(q_1 \cdot x + q_2 \cdot y + q_3 \cdot z)} \< 0 | T \{ j_\mathrm{em}^\mu(x) j_\mathrm{em}^\nu(y) j_\mathrm{em}^\lambda(z) j_\mathrm{em}^\sigma(0) \} | 0 \>  \, ,
\end{align}
which due to gauge invariance needs to satisfy the Ward--Takahashi (WT) identities
\begin{align}
	\{q_1^\mu, q_2^\nu, q_3^\lambda, q_4^\sigma\} \Pi_{\mu\nu\lambda\sigma}(q_1,q_2,q_3) = 0 \, ,
\end{align}
where $q_4 = q_1 + q_2 + q_3$. Based on the recipe by Bardeen, Tung~\cite{Bardeen:1969aw}, and Tarrach~\cite{Tarrach:1975tu} for generic photon amplitudes, a tensor decomposition
\begin{align}
	\label{eq:HLbLTensorDecomposition}
	\Pi^{\mu\nu\lambda\sigma}(q_1,q_2,q_3) &= \sum_{i=1}^{54}
        T_i^{\mu\nu\lambda\sigma} \Pi_i(s,t,u) ,  
\end{align}
into a redundant set of 54 Lorentz structures was derived in Refs.~\cite{Colangelo:2015ama,Colangelo:2017fiz}, in such a way that the structures $T_i^{\mu\nu\lambda\sigma}$ individually satisfy the WT identities, while at the same time the scalar functions $\Pi_i$ are free of kinematic singularities and zeros. The Mandelstam variables are defined by $s = (q_1+q_2)^2$, $t = (q_1 + q_3)^2$, $u = (q_2 + q_3)^2$. Thanks to the absence of kinematic singularities, a master formula for the HLbL contribution to $a_\mu$ holds directly in terms of the hadronic scalar functions $\Pi_i$,
\begin{align}
	\label{eq:Masterformula}
	a_\mu^\mathrm{HLbL} &= \frac{2 \alpha^3}{3 \pi^2} \int_0^\infty
        dQ_1 \int_0^\infty dQ_2 \int_{-1}^1 d\tau \sqrt{1-\tau^2} Q_1^3
        Q_2^3 \sum_{i=1}^{12} T_i(Q_1,Q_2,\tau) \bar \Pi_i(Q_1,Q_2,\tau) \, ,
\end{align}
where $T_i$ are known integration kernels and only 12 independent linear combinations $\bar \Pi_i$ of the 
$\Pi_i$ contribute in the reduced kinematics
\begin{align}
	\label{eq:ReducedKinematics}
	s &= - Q_3^2 \,, \quad t = - Q_2^2 \,, \quad u = - Q_1^2 \,, \nn
	q_1^2 &= -Q_1^2 \,, \quad q_2^2 = - Q_2^2 \,, \quad q_3^2 = - Q_3^2 = - Q_1^2 - 2 Q_1 Q_2 \tau - Q_2^2 \,, \quad q_4^2 = 0 \, .
\end{align}
Further, the decomposition~\eqref{eq:HLbLTensorDecomposition} allowed us to set up a dispersive framework for HLbL in four-point kinematics, exploiting analyticity and unitarity similarly to Eq.~\eqref{eq:amuHVP}. The dispersive approach enabled the evaluation of the dominant contributions to HLbL with controlled and much reduced uncertainties, in particular of the pseudoscalar-pole and two-pion contributions, leading to~\cite{Melnikov:2003xd,Masjuan:2017tvw,Colangelo:2017fiz,Hoferichter:2018kwz,Gerardin:2019vio,Bijnens:2019ghy,Colangelo:2019uex,Pauk:2014rta,Danilkin:2016hnh,Jegerlehner:2017gek,Knecht:2018sci,Eichmann:2019bqf,Roig:2019reh}
\begin{align}
	\label{eq:amuHLbL}
	a_\mu^\mathrm{HLbL, WP} = 92(19) \times 10^{-11} \, .
\end{align}
The uncertainty in Eq.~\eqref{eq:amuHLbL} is dominated on the one hand by the contributions of scalar, axial-vector, and tensor resonances in the $(1\text{--}2)\GeV$ range, which were not computed dispersively yet but estimated using hadronic models, on the other hand by the matching to short-distance constraints (SDCs) that follow from the operator-product expansion (OPE). Therefore, since the publication of the WP, efforts were directed towards an improved evaluation of these sub-dominant contributions.

Scalar resonances beyond the $f_0(500)$ were evaluated within the dispersive framework in Ref.~\cite{Danilkin:2021icn}, based on a coupled-channel treatment of $\gamma^*\gamma^*\to\pi\pi/\bar KK$, leading to only a small increase of the $S$-wave contribution from Ref.~\cite{Colangelo:2017fiz}. A dispersive analysis of kaon vector form factors~\cite{Stamen:2022uqh} provided a tiny value for the kaon box, in line with previous estimates~\cite{Eichmann:2019bqf,Aoyama:2020ynm,Miramontes:2021exi}. Axial-vector contributions and their interplay with SDCs have been studied within holographic QCD models~\cite{Leutgeb:2019gbz,Cappiello:2019hwh,Leutgeb:2022lqw}. The inclusion of axial-vector resonances within the dispersive framework became only possible with a modified tensor decomposition discussed in Ref.~\cite{Colangelo:2021nkr}, with crucial input required for the axial-vector transition form factors. 
Asymptotic constraints on these form factors were worked out in Ref.~\cite{Hoferichter:2020lap} and the available experimental constraints analyzed within vector-meson-dominance-inspired parameterizations in Refs.~\cite{Zanke:2021wiq,Hoferichter:2023tgp}.  

For the inclusion of two-pion rescattering beyond the $S$-wave, the sub-process $\gamma^*\gamma^*\to\pi\pi$ was reconstructed dispersively up to $D$-waves in Refs.~\cite{Hoferichter:2019nlq,Danilkin:2019opj}, but so far the inclusion in the dispersive framework for HLbL has not been possible due to the appearance of kinematic singularities that ultimately can be traced back to the redundancy in the decomposition~\eqref{eq:HLbLTensorDecomposition}. In order to include higher partial waves as well as tensor-meson resonances, the dispersive framework itself needs to be extended. A solution to this problem was proposed in Ref.~\cite{Ludtke:2023hvz}, establishing dispersion relations not for four-point kinematics, but directly in the kinematic limit~\eqref{eq:ReducedKinematics}. This modified framework requires the reconstruction of further sub-processes~\cite{Ludtke:2023vgd}, but in combination with the established dispersive framework it promises to enable a complete dispersive evaluation of HLbL.

The SDCs on HLbL have recently seen further improvement as well: perturbative and non-perturbative corrections to the OPEs in the different asymptotic limits were studied in Refs.~\cite{Bijnens:2020xnl,Bijnens:2021jqo,Bijnens:2022itw} and were used for an improved matching based on a model of excited pseudoscalars in Ref.~\cite{Colangelo:2021nkr}, see also Refs.~\cite{Knecht:2020xyr,Masjuan:2020jsf,Ludtke:2020moa} for the implementation of the SDCs. Work is in progress to combine all these developments, together with improvements of the $\eta$, $\eta'$ pole contributions~\cite{Holz:2015tcg,Holz:2022hwz,Alexandrou:2022qyf,Gerardin:2023naa},  into a full data-driven evaluation of $a_\mu^\text{HLbL}$.

Since the WP publication, the HLbL contribution has also been evaluated within lattice QCD with competitive uncertainties. The results
\begin{align}
	a_\mu^\text{HLbL, lattice} &= 109.6(15.9) \times 10^{-11} \quad \text{\cite{Chao:2021tvp,Chao:2022xzg}} \, , \nn
	a_\mu^\text{HLbL, lattice} &= 124.7(14.9) \times 10^{-11} \quad  \text{\cite{Blum:2023vlm}},
\end{align}
are compatible with the phenomenological WP value~\eqref{eq:amuHLbL}, but point to a slightly larger central value.
In order to meet the final experimental precision goal, the uncertainties in HLbL should be further reduced to the level of about $10\%$, which seems feasible for both phenomenological and lattice-QCD evaluations.
However, all these improvement in the HLbL evaluation will only have a real impact once the tensions in the evaluations of HVP are resolved, to which we turn next.

\section{Hadronic vacuum polarization}
\label{sec:HVP}

The discrepancy between the experimental value for $a_\mu$~\eqref{eq:amuExp} and the SM evaluation~\eqref{eq:amuSM} is reduced to only $1.5\sigma$ if the evaluation of HVP is replaced by~\cite{Borsanyi:2020mff}
\begin{align}
	a_\mu^\text{HVP, LO, BMWc} = 7\,075(55) \times 10^{-11} \, .
\end{align}
However, this number is in $2.1\sigma$ tension with the WP evaluation based on $e^+e^-$ cross-section data. The resolution of this tension is crucial in order to update the WP prediction and to reach a single competitive SM prediction for $a_\mu$~\cite{Colangelo:2022jxc}. The current puzzling situation has triggered intense scrutiny of both the lattice and dispersive evaluations. So-called window quantities, obtained by introducing weight functions in the Euclidean-time integral of the coordinate-space representation of HVP~\cite{RBC:2018dos}, have proved useful, as the intermediate window is much less affected by lattice systematics than the entire HVP contribution to $a_\mu$. The BMWc value for this quantity is in $3.7\sigma$ tension with the cross-section data~\cite{Colangelo:2022vok} and several lattice collaborations have now confirmed this result~\cite{Ce:2022kxy,ExtendedTwistedMass:2022jpw,FermilabLatticeHPQCD:2023jof,Blum:2023qou}.

At the current level of precision, the lattice calculations need to account for both strong isospin breaking as well as QED corrections to the two-point correlator. This part of the lattice calculation has been compared to a phenomenological evaluation of isospin-breaking effects in Refs.~\cite{Hoferichter:2023sli,Hoferichter:2023bjm,Colangelo:2022prz,Colangelo:2021moe,Hoferichter:2022iqe}, which found that the lattice calculation of these contributions cannot be the source of the discrepancy. Whether radiative corrections on the experimental side could provide an explanation of the puzzle is the subject of ongoing research~\cite{Campanario:2019mjh,Colangelo:2022lzg,Ignatov:2022iou,JMPhDThesis,Abbiendi:2022liz}.

Localizing the source of the discrepancy to a certain energy range in the hadronic cross sections is an ill-posed inverse problem~\cite{Colangelo:2022vok}, but the information from the window quantities as well as the constraints on the hadronic running of $\alpha$ imply that the differences mainly come from the region below $\approx 2\GeV$~\cite{Passera:2008jk,Crivellin:2020zul,Keshavarzi:2020bfy,Malaescu:2020zuc,Colangelo:2020lcg,Ce:2022eix}. In the low-energy region, the two-pion channel completely dominates and modifications of the cross-section data can be confronted with the constraints of analyticity and unitarity on the pion vector form factor (VFF)~\cite{Colangelo:2020lcg}. We are using the representation for the VFF~\cite{Colangelo:2018mtw}
\begin{align}
	\label{eq:VFF}
	F_\pi^V(s) &= \Omega_1^1(s) \times G_\omega(s) \times G_\mathrm{in}^N(s) \, ,
\end{align}
where
\begin{align}
	\Omega_1^1(s) = \exp\left\{ \frac{s}{\pi} \int_{4\mpi^2}^\infty ds^\prime \frac{\delta_1^1(s^\prime)}{s^\prime(s^\prime-s)} \right\}
\end{align}
denotes the Omn\`es function with the elastic $\pi\pi$-scattering $P$-wave phase shift $\delta_1^1(s)$ as input~\cite{Ananthanarayan:2000ht,Caprini:2011ky}, the second factor accounts for the resonantly enhanced isospin-breaking $\rho$--$\omega$ interference effect
\begin{align}
	G_\omega(s) &= 1 + \frac{s}{\pi} \int_{9\mpi^2}^\infty ds^\prime \frac{\Im g_\omega(s^\prime)}{s^\prime(s^\prime-s)} \left( \frac{1 - \frac{9\mpi^2}{s^\prime}}{1 - \frac{9\mpi^2}{M_\omega^2}} \right)^4 \, , \quad g_\omega(s) = 1 + \epsilon_\omega \frac{s}{(M_\omega - \frac{i}{2} \Gamma_\omega)^2 - s} \, ,
\end{align}
and further inelastic contributions are parametrized by a conformal polynomial $G_\mathrm{in}^N(s)$ with a cut starting at the $\pi^0\omega$ threshold. Although this dispersive representation only depends on a few parameters, there is enough freedom to describe all major experiments individually---in particular, the constraints of analyticity and unitarity do not resolve the tension between BaBar~\cite{Lees:2012cj} and KLOE~\cite{Anastasi:2017eio}, see Ref.~\cite{Colangelo:2018mtw}. Similarly, even modifications of the cross-section data well beyond the BaBar--KLOE tension can be accommodated by the dispersive constraints, with rather uniform shifts in the two-pion cross section leading to a correlated shift in the pion charge radius~\cite{Colangelo:2020lcg}, which potentially could provide an independent cross check if an improved lattice determination of the pion charge radius became available~\cite{Feng:2019geu,Wang:2020nbf}.

Exactly such a shift in the cross-section data is indeed realized in the recent measurements by \mbox{CMD-3}~\cite{CMD-3:2023alj}. In Table~\ref{tab:FitResults}, we present updated results for the fit of the dispersive representation~\eqref{eq:VFF} to the major experiments, including CMD-3. Compared to Refs.~\cite{Colangelo:2018mtw,Colangelo:2022prz}, we update the input for the $\omega$ width to the $3\pi$ result $\Gamma_\omega = 8.71(6)\MeV$~\cite{Hoferichter:2023bjm}.
We find a $p$-value of $20\%$ for the fit to CMD-3: the data are compatible with the dispersive constraints. Our dispersive representation also allows us to quantify the tension to the other experiments for the full energy range up to $1\GeV$, shown in Table~\ref{tab:Tensions}. As noted in Ref.~\cite{Colangelo:2022prz}, SND20~\cite{SND:2020nwa} is the only experiment that cannot be fit with a good $p$-value, while BaBar, KLOE, BESIII~\cite{Ablikim:2015orh}, SND06~\cite{Achasov:2006vp}, and CMD-2~\cite{Akhmetshin:2006bx} do permit acceptable fits.  

\begin{figure}[t]
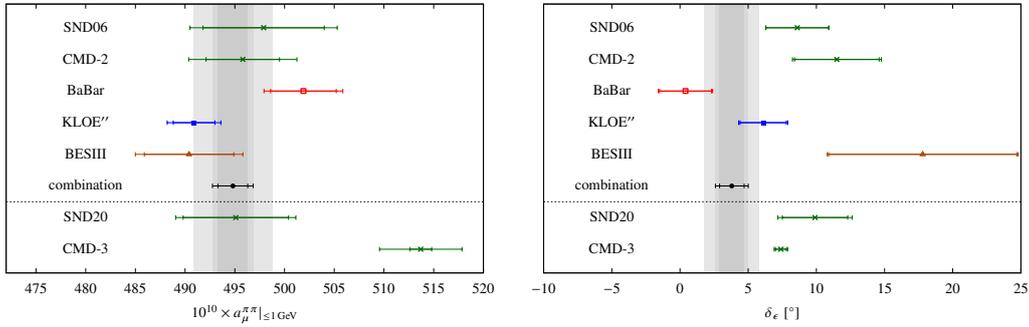
 
	\centering\small
	\scalebox{0.55}{\input{plots/amu}}
	\scalebox{0.55}{\input{plots/epsphase}}
	\caption{Left: Results for $a_\mu^{\pi\pi}$ in the energy range $\le1\GeV$. Right: Results for the phase of the $\rho$--$\omega$ mixing parameter, $\delta_\epsilon$. The smaller error bars refer to the fit uncertainties, inflated by $\sqrt{\chi^2/\mathrm{dof}}$, the larger error bars to the total uncertainties. The gray bands correspond to the combined fit to NA7 and all $e^+e^-$ data sets apart from SND20 and CMD-3, with the largest band including the additional systematic effect due to the BaBar--KLOE tension.}
	\label{fig:Plots}
\end{figure}

The discrepancies among the different $e^+e^-$ experiments are shown in Fig.~\ref{fig:Plots}. For $a_\mu$ itself, the discrepancy between CMD-3 and the combination of the other experiments by far exceeds the BaBar--KLOE tension or the one between BMWc and the WP, amounting to $5\sigma$ for the HVP integral up to $1\GeV$ and even more around the $\rho$ peak or in the intermediate window. 
Further tensions are visible directly in the fit parameters, e.g., the complex phase 
$\delta_\epsilon$ of the $\rho$--$\omega$ mixing parameter $\epsilon_\omega$, an observable generated by radiative channels such as $\rho\to\pi^0\gamma\to\omega$~\cite{Colangelo:2022prz}, differs widely among the experiments.

\begin{table}[t]
	\centering
	\scriptsize
	\scalebox{0.95}{\begin{tabularx}{1.05\textwidth}{l l l l l l}
	\toprule
	 	& $\chi^2/\mathrm{dof}$ & $p$-value & $M_\omega$ [MeV] & $10^3 \times \Re\epsilon_\omega$ & $\delta_\epsilon$ [${}^\circ$] \\
	\midrule
	SND06					&	$1.09$	&	$33\%$			&	$782.12(33)(2)$	&	$2.03(5)(2)$	&	$8.6(2.3)(0.3)$		 \\
	CMD-2					&	$1.01$	&	$46\%$			&	$782.65(33)(4)$	&	$1.90(6)(3)$	&	$11.5(3.1)(1.0)$		 \\
	BaBar					&	$1.17$	&	$3.0\%$			&	$781.89(18)(4)$	&	$2.06(4)(2)$	&	$0.4(1.9)(0.6)$		 \\
	KLOE$''$					&	$1.13$	&	$11\%$			&	$782.45(24)(5)$	&	$1.96(4)(2)$	&	$6.1(1.7)(0.6)$		 \\
	BESIII					&	$1.01$	&	$45\%$			&	$783.07(61)(2)$	&	$2.03(19)(7)$	&	$17.8(6.9)(1.2)$		 \\
	SND20					&	$1.88$	&	$3.8\times10^{-3}$	&	$782.34(28)(6)$	&	$2.07(5)(2)$	&	$9.9(2.4)(1.3)$		 \\
	CMD-3					&	$1.09$	&	$20\%$			&	$782.33(6)(3)$		&	$2.08(1)(2)$	&	$7.4(4)(3)$	 \\
	\midrule
	Combination					&	$1.21$	&	$1.4\times10^{-4}$	&	$782.07(12)(5)(8)$	&	$1.99(2)(2)(0)$	&	$3.8(0.9)(0.8)(1.6)$		 \\
	\midrule
	\midrule
	\end{tabularx}}
	\scalebox{0.95}{\begin{tabularx}{1.05\textwidth}{l l l l l l}
	 \hfill{}$10^{10} \times{}$	& $a_\mu^{\pi\pi} \big|_{[0.60,0.88]\GeV}$ & $a_\mu^{\pi\pi} \big|_{\le 1\GeV}$ & SD window  & int window & LD window \\
	\midrule
	SND06					&	$366.2(4.9)(2.7)$	&	$497.9(6.1)(4.2)$	&	$13.9(2)(1)$		&	$139.6(1.8)(1.0)$	&	$344.4(4.1)(3.1)$ \\
	CMD-2					&	$365.7(2.9)(2.0)$	&	$495.8(3.7)(4.0)$	&	$13.9(1)(1)$		&	$139.4(1.0)(0.8)$	&	$342.6(2.5)(3.1)$ \\
	BaBar					&	$368.5(2.7)(1.9)$	&	$501.9(3.3)(2.2)$	&	$14.0(1)(1)$		&	$140.6(1.0)(0.7)$	&	$347.4(2.2)(1.4)$ \\
	KLOE$''$					&	$359.8(1.6)(1.0)$	&	$490.9(2.1)(1.7)$	&	$13.6(1)(0)$		&	$137.1(0.6)(0.4)$	&	$340.2(1.4)(1.2)$ \\
	BESIII					&	$361.4(3.6)(1.7)$	&	$490.4(4.5)(3.0)$	&	$13.7(1)(0)$		&	$137.8(1.3)(0.4)$	&	$338.9(3.1)(2.6)$ \\
	SND20					&	$364.4(4.2)(2.2)$	&	$495.1(5.3)(2.9)$	&	$13.8(2)(0)$		&	$139.2(1.5)(0.4)$	&	$342.0(3.7)(2.4)$ \\
	CMD-3					&	$378.7(0.8)(2.9)$	&	$513.7(1.1)(4.0)$	&	$14.3(0)(1)$		&	$144.0(0.3)(1.1)$	&	$355.4(0.7)(2.7)$\\
	\midrule
	Combination					&	$363.0(1.2)(0.8)(2.7)$	&	$494.8(1.5)(1.4)(3.4)$	&	$13.7(0)(0)(1)$	&	$138.3(0.4)(0.3)(1.1)$	&	$342.7(1.0)(1.1)(2.2)$		 \\
	\bottomrule
	\end{tabularx}}%
	\caption{The row ``combination'' includes the space-like data from NA7~\cite{Amendolia:1986wj} and all $e^+e^-$ data sets apart from SND20 and CMD-3. The first error is the fit error, including $\chi^2$ inflation, the second error includes all systematics.
	In the combined fit, the third error is the systematic effect due to the BaBar--KLOE tension, according to the WP prescription.}
	\label{tab:FitResults}
\end{table}

\begin{table}[t]
	\centering
	\footnotesize
	\begin{tabularx}{9cm}{X l l l}
	\toprule
	Discrepancy	& $a_\mu^{\pi\pi} \big|_{[0.60,0.88]\GeV}$ & $a_\mu^{\pi\pi} \big|_{\le 1\GeV}$  & int window \\
	\midrule
	SND06					&	$1.8\sigma$	&	$1.7\sigma$ 	&	$1.7\sigma$ \\
	CMD-2					&	$2.3\sigma$	&	$2.0\sigma$ 	&	$2.1\sigma$ \\
	BaBar					&	$3.3\sigma$	&	$2.9\sigma$ 	&	$3.1\sigma$ \\
	KLOE$''$					&	$5.6\sigma$	&	$4.8\sigma$ 	&	$5.4\sigma$ \\
	BESIII					&	$3.0\sigma$	&	$2.8\sigma$ 	&	$3.1\sigma$ \\
	SND20					&	$2.2\sigma$	&	$2.1\sigma$ 	&	$2.2\sigma$ \\
	\midrule
	Combination					&	$4.2\sigma$ [$6.1\sigma$]	&	$3.7\sigma$ [$5.0\sigma$] 	&	$3.8\sigma$ [$5.7\sigma$] \\
	\bottomrule
	\end{tabularx}
	\caption{Significance of the discrepancies between fits to CMD-3 and the other experiments, taking into account the correlations due to the systematics in the dispersive representation, as well as the $\chi^2$ inflation of the fit errors. For the combined fit, the discrepancies in square brackets exclude the systematic effect due to the BaBar--KLOE tension.}
	\label{tab:Tensions}
\end{table}

\section{Conclusions}

The evaluation of the hadronic contributions to $a_\mu$ has been the subject of intense research efforts, both using lattice QCD and data-driven methods. While recent work on HLbL promises to reach the precision goal set by the Fermilab experiment, the interpretation of the SM prediction is currently complicated by the presence of a multitude of puzzles in the HVP contribution: the disagreement between lattice QCD and hadronic cross sections on the one hand, but also a new discrepancy between CMD-3 and all other $e^+e^-\to\pi^+\pi^-$ experiments on the other. We presented updated fit results of our dispersive representation of the pion vector form factor to the $e^+e^-\to\pi^+\pi^-$ data sets. The fit to the CMD-3 data did not reveal any conflict with the dispersive constraints, yet the discrepancies to the other experiments are substantial: when compared to the combination they amount to $5\sigma$ for the entire energy range below $1\GeV$, even more for some partial quantities, see Table~\ref{tab:Tensions}. Forthcoming results from ongoing $e^+e^-\to\pi^+\pi^-$ analyses, a reinvestigation of radiative corrections, and further lattice-QCD computations scrutinizing the BMWc result for the full HVP contribution will be indispensable to understand the current puzzling situation.


\acknowledgments

We gratefully acknowledge financial support by the Swiss National Science Foundation (Project Nos.\ 200020\_175791, PCEFP2\_181117, and PCEFP2\_194272).


\phantomsection
\addcontentsline{toc}{section}{\numberline{}References}
\bibliographystyle{apsrev4-1_mod}
\bibliography{amu}

\begin{thebibliography}{107}%
\makeatletter
\providecommand \@ifxundefined [1]{%
 \@ifx{#1\undefined}
}%
\providecommand \@ifnum [1]{%
 \ifnum #1\expandafter \@firstoftwo
 \else \expandafter \@secondoftwo
 \fi
}%
\providecommand \@ifx [1]{%
 \ifx #1\expandafter \@firstoftwo
 \else \expandafter \@secondoftwo
 \fi
}%
\providecommand \natexlab [1]{#1}%
\providecommand \enquote  [1]{``#1''}%
\providecommand \bibnamefont  [1]{#1}%
\providecommand \bibfnamefont [1]{#1}%
\providecommand \citenamefont [1]{#1}%
\providecommand \href@noop [0]{\@secondoftwo}%
\providecommand \href [0]{\begingroup \@sanitize@url \@href}%
\providecommand \@href[1]{\@@startlink{#1}\@@href}%
\providecommand \@@href[1]{\endgroup#1\@@endlink}%
\providecommand \@sanitize@url [0]{\catcode `\\12\catcode `\$12\catcode
  `\&12\catcode `\#12\catcode `\^12\catcode `\_12\catcode `\%12\relax}%
\providecommand \@@startlink[1]{}%
\providecommand \@@endlink[0]{}%
\providecommand \url  [0]{\begingroup\@sanitize@url \@url }%
\providecommand \@url [1]{\endgroup\@href {#1}{\urlprefix }}%
\providecommand \urlprefix  [0]{URL }%
\providecommand \Eprint [0]{\href }%
\providecommand \doibase [0]{http://dx.doi.org/}%
\providecommand \selectlanguage [0]{\@gobble}%
\providecommand \bibinfo  [0]{\@secondoftwo}%
\providecommand \bibfield  [0]{\@secondoftwo}%
\providecommand \translation [1]{[#1]}%
\providecommand \BibitemOpen [0]{}%
\providecommand \bibitemStop [0]{}%
\providecommand \bibitemNoStop [0]{.\EOS\space}%
\providecommand \EOS [0]{\spacefactor3000\relax}%
\providecommand \BibitemShut  [1]{\csname bibitem#1\endcsname}%
\let\auto@bib@innerbib\@empty
\bibitem [{\citenamefont {Abi}\ \emph {et~al.}(2021)\citenamefont {Abi} \emph
  {et~al.}}]{Muong-2:2021ojo}%
  \BibitemOpen
  \bibfield  {author} {\bibinfo {author} {\bibfnamefont {B.}~\bibnamefont
  {Abi}}  \emph {et~al.} (\bibinfo {collaboration} {Muon $g-2$}),\ }\href
  {\doibase 10.1103/PhysRevLett.126.141801} {\bibfield  {journal} {\bibinfo
  {journal} {Phys. Rev. Lett.}\ }\textbf {\bibinfo {volume} {126}},\ \bibinfo
  {pages} {141801} (\bibinfo {year} {2021})},\ \Eprint
  {http://arxiv.org/abs/2104.03281} {arXiv:2104.03281 [hep-ex]}\BibitemShut
  {NoStop}%
\bibitem [{\citenamefont {Albahri}\ \emph
  {et~al.}(2021{\natexlab{a}})\citenamefont {Albahri} \emph
  {et~al.}}]{Muong-2:2021ovs}%
  \BibitemOpen
  \bibfield  {author} {\bibinfo {author} {\bibfnamefont {T.}~\bibnamefont
  {Albahri}}  \emph {et~al.} (\bibinfo {collaboration} {Muon $g-2$}),\ }\href
  {\doibase 10.1103/PhysRevA.103.042208} {\bibfield  {journal} {\bibinfo
  {journal} {Phys. Rev. A}\ }\textbf {\bibinfo {volume} {103}},\ \bibinfo
  {pages} {042208} (\bibinfo {year} {2021}{\natexlab{a}})},\ \Eprint
  {http://arxiv.org/abs/2104.03201} {arXiv:2104.03201 [hep-ex]}\BibitemShut
  {NoStop}%
\bibitem [{\citenamefont {Albahri}\ \emph
  {et~al.}(2021{\natexlab{b}})\citenamefont {Albahri} \emph
  {et~al.}}]{Muong-2:2021xzz}%
  \BibitemOpen
  \bibfield  {author} {\bibinfo {author} {\bibfnamefont {T.}~\bibnamefont
  {Albahri}}  \emph {et~al.} (\bibinfo {collaboration} {Muon $g-2$}),\ }\href
  {\doibase 10.1103/PhysRevAccelBeams.24.044002} {\bibfield  {journal}
  {\bibinfo  {journal} {Phys. Rev. Accel. Beams}\ }\textbf {\bibinfo {volume}
  {24}},\ \bibinfo {pages} {044002} (\bibinfo {year} {2021}{\natexlab{b}})},\
  \Eprint {http://arxiv.org/abs/2104.03240} {arXiv:2104.03240
  [physics.acc-ph]}\BibitemShut {NoStop}%
\bibitem [{\citenamefont {Albahri}\ \emph
  {et~al.}(2021{\natexlab{c}})\citenamefont {Albahri} \emph
  {et~al.}}]{Muong-2:2021vma}%
  \BibitemOpen
  \bibfield  {author} {\bibinfo {author} {\bibfnamefont {T.}~\bibnamefont
  {Albahri}}  \emph {et~al.} (\bibinfo {collaboration} {Muon $g-2$}),\ }\href
  {\doibase 10.1103/PhysRevD.103.072002} {\bibfield  {journal} {\bibinfo
  {journal} {Phys. Rev. D}\ }\textbf {\bibinfo {volume} {103}},\ \bibinfo
  {pages} {072002} (\bibinfo {year} {2021}{\natexlab{c}})},\ \Eprint
  {http://arxiv.org/abs/2104.03247} {arXiv:2104.03247 [hep-ex]}\BibitemShut
  {NoStop}%
\bibitem [{\citenamefont {Bennett}\ \emph {et~al.}(2006)\citenamefont {Bennett}
  \emph {et~al.}}]{Muong-2:2006rrc}%
  \BibitemOpen
  \bibfield  {author} {\bibinfo {author} {\bibfnamefont {G.~W.}\ \bibnamefont
  {Bennett}}  \emph {et~al.} (\bibinfo {collaboration} {Muon $g-2$}),\ }\href
  {\doibase 10.1103/PhysRevD.73.072003} {\bibfield  {journal} {\bibinfo
  {journal} {Phys. Rev. D}\ }\textbf {\bibinfo {volume} {73}},\ \bibinfo
  {pages} {072003} (\bibinfo {year} {2006})},\ \Eprint
  {http://arxiv.org/abs/hep-ex/0602035} {arXiv:hep-ex/0602035}\BibitemShut
  {NoStop}%
\bibitem [{\citenamefont {Aoyama}\ \emph {et~al.}(2012)\citenamefont {Aoyama},
  \citenamefont {Hayakawa}, \citenamefont {Kinoshita},\ and\ \citenamefont
  {Nio}}]{Aoyama:2012wk}%
  \BibitemOpen
  \bibfield  {author} {\bibinfo {author} {\bibfnamefont {T.}~\bibnamefont
  {Aoyama}}, \bibinfo {author} {\bibfnamefont {M.}~\bibnamefont {Hayakawa}},
  \bibinfo {author} {\bibfnamefont {T.}~\bibnamefont {Kinoshita}}, and \bibinfo
  {author} {\bibfnamefont {M.}~\bibnamefont {Nio}},\ }\href {\doibase
  10.1103/PhysRevLett.109.111808} {\bibfield  {journal} {\bibinfo  {journal}
  {Phys. Rev. Lett.}\ }\textbf {\bibinfo {volume} {109}},\ \bibinfo {pages}
  {111808} (\bibinfo {year} {2012})},\ \Eprint {http://arxiv.org/abs/1205.5370}
  {arXiv:1205.5370 [hep-ph]}\BibitemShut {NoStop}%
\bibitem [{\citenamefont {Aoyama}\ \emph {et~al.}(2019)\citenamefont {Aoyama},
  \citenamefont {Kinoshita},\ and\ \citenamefont {Nio}}]{Aoyama:2019ryr}%
  \BibitemOpen
  \bibfield  {author} {\bibinfo {author} {\bibfnamefont {T.}~\bibnamefont
  {Aoyama}}, \bibinfo {author} {\bibfnamefont {T.}~\bibnamefont {Kinoshita}},
  and \bibinfo {author} {\bibfnamefont {M.}~\bibnamefont {Nio}},\ }\href
  {\doibase 10.3390/atoms7010028} {\bibfield  {journal} {\bibinfo  {journal}
  {Atoms}\ }\textbf {\bibinfo {volume} {7}},\ \bibinfo {pages} {28} (\bibinfo
  {year} {2019})}\BibitemShut {NoStop}%
\bibitem [{\citenamefont {Czarnecki}\ \emph {et~al.}(2003)\citenamefont
  {Czarnecki}, \citenamefont {Marciano},\ and\ \citenamefont
  {Vainshtein}}]{Czarnecki:2002nt}%
  \BibitemOpen
  \bibfield  {author} {\bibinfo {author} {\bibfnamefont {A.}~\bibnamefont
  {Czarnecki}}, \bibinfo {author} {\bibfnamefont {W.~J.}\ \bibnamefont
  {Marciano}}, and \bibinfo {author} {\bibfnamefont {A.}~\bibnamefont
  {Vainshtein}},\ }\href {\doibase 10.1103/PhysRevD.67.073006} {\bibfield
  {journal} {\bibinfo  {journal} {Phys. Rev. D}\ }\textbf {\bibinfo {volume}
  {67}},\ \bibinfo {pages} {073006} (\bibinfo {year} {2003})},\ \bibinfo {note}
  {[Erratum: Phys. Rev. D {\bf 73}, 119901 (2006)]},\ \Eprint
  {http://arxiv.org/abs/hep-ph/0212229} {arXiv:hep-ph/0212229}\BibitemShut
  {NoStop}%
\bibitem [{\citenamefont {Gnendiger}\ \emph {et~al.}(2013)\citenamefont
  {Gnendiger}, \citenamefont {St\"ockinger},\ and\ \citenamefont
  {St\"ockinger-Kim}}]{Gnendiger:2013pva}%
  \BibitemOpen
  \bibfield  {author} {\bibinfo {author} {\bibfnamefont {C.}~\bibnamefont
  {Gnendiger}}, \bibinfo {author} {\bibfnamefont {D.}~\bibnamefont
  {St\"ockinger}}, and \bibinfo {author} {\bibfnamefont {H.}~\bibnamefont
  {St\"ockinger-Kim}},\ }\href {\doibase 10.1103/PhysRevD.88.053005} {\bibfield
   {journal} {\bibinfo  {journal} {Phys. Rev. D}\ }\textbf {\bibinfo {volume}
  {88}},\ \bibinfo {pages} {053005} (\bibinfo {year} {2013})},\ \Eprint
  {http://arxiv.org/abs/1306.5546} {arXiv:1306.5546 [hep-ph]}\BibitemShut
  {NoStop}%
\bibitem [{\citenamefont {Davier}\ \emph {et~al.}(2017)\citenamefont {Davier},
  \citenamefont {Hoecker}, \citenamefont {Malaescu},\ and\ \citenamefont
  {Zhang}}]{Davier:2017zfy}%
  \BibitemOpen
  \bibfield  {author} {\bibinfo {author} {\bibfnamefont {M.}~\bibnamefont
  {Davier}}, \bibinfo {author} {\bibfnamefont {A.}~\bibnamefont {Hoecker}},
  \bibinfo {author} {\bibfnamefont {B.}~\bibnamefont {Malaescu}}, and \bibinfo
  {author} {\bibfnamefont {Z.}~\bibnamefont {Zhang}},\ }\href {\doibase
  10.1140/epjc/s10052-017-5161-6} {\bibfield  {journal} {\bibinfo  {journal}
  {Eur. Phys. J. C}\ }\textbf {\bibinfo {volume} {77}},\ \bibinfo {pages} {827}
  (\bibinfo {year} {2017})},\ \Eprint {http://arxiv.org/abs/1706.09436}
  {arXiv:1706.09436 [hep-ph]}\BibitemShut {NoStop}%
\bibitem [{\citenamefont {Keshavarzi}\ \emph {et~al.}(2018)\citenamefont
  {Keshavarzi}, \citenamefont {Nomura},\ and\ \citenamefont
  {Teubner}}]{Keshavarzi:2018mgv}%
  \BibitemOpen
  \bibfield  {author} {\bibinfo {author} {\bibfnamefont {A.}~\bibnamefont
  {Keshavarzi}}, \bibinfo {author} {\bibfnamefont {D.}~\bibnamefont {Nomura}},
  and \bibinfo {author} {\bibfnamefont {T.}~\bibnamefont {Teubner}},\ }\href
  {\doibase 10.1103/PhysRevD.97.114025} {\bibfield  {journal} {\bibinfo
  {journal} {Phys. Rev. D}\ }\textbf {\bibinfo {volume} {97}},\ \bibinfo
  {pages} {114025} (\bibinfo {year} {2018})},\ \Eprint
  {http://arxiv.org/abs/1802.02995} {arXiv:1802.02995 [hep-ph]}\BibitemShut
  {NoStop}%
\bibitem [{\citenamefont {Colangelo}\ \emph {et~al.}(2019)\citenamefont
  {Colangelo}, \citenamefont {Hoferichter},\ and\ \citenamefont
  {Stoffer}}]{Colangelo:2018mtw}%
  \BibitemOpen
  \bibfield  {author} {\bibinfo {author} {\bibfnamefont {G.}~\bibnamefont
  {Colangelo}}, \bibinfo {author} {\bibfnamefont {M.}~\bibnamefont
  {Hoferichter}}, and \bibinfo {author} {\bibfnamefont {P.}~\bibnamefont
  {Stoffer}},\ }\href {\doibase 10.1007/JHEP02(2019)006} {\bibfield  {journal}
  {\bibinfo  {journal} {JHEP}\ }\textbf {\bibinfo {volume} {02}},\ \bibinfo
  {pages} {006} (\bibinfo {year} {2019})},\ \Eprint
  {http://arxiv.org/abs/1810.00007} {arXiv:1810.00007 [hep-ph]}\BibitemShut
  {NoStop}%
\bibitem [{\citenamefont {Hoferichter}\ \emph {et~al.}(2019)\citenamefont
  {Hoferichter}, \citenamefont {Hoid},\ and\ \citenamefont
  {Kubis}}]{Hoferichter:2019gzf}%
  \BibitemOpen
  \bibfield  {author} {\bibinfo {author} {\bibfnamefont {M.}~\bibnamefont
  {Hoferichter}}, \bibinfo {author} {\bibfnamefont {B.-L.}\ \bibnamefont
  {Hoid}}, and \bibinfo {author} {\bibfnamefont {B.}~\bibnamefont {Kubis}},\
  }\href {\doibase 10.1007/JHEP08(2019)137} {\bibfield  {journal} {\bibinfo
  {journal} {JHEP}\ }\textbf {\bibinfo {volume} {08}},\ \bibinfo {pages} {137}
  (\bibinfo {year} {2019})},\ \Eprint {http://arxiv.org/abs/1907.01556}
  {arXiv:1907.01556 [hep-ph]}\BibitemShut {NoStop}%
\bibitem [{\citenamefont {Davier}\ \emph {et~al.}(2020)\citenamefont {Davier},
  \citenamefont {Hoecker}, \citenamefont {Malaescu},\ and\ \citenamefont
  {Zhang}}]{Davier:2019can}%
  \BibitemOpen
  \bibfield  {author} {\bibinfo {author} {\bibfnamefont {M.}~\bibnamefont
  {Davier}}, \bibinfo {author} {\bibfnamefont {A.}~\bibnamefont {Hoecker}},
  \bibinfo {author} {\bibfnamefont {B.}~\bibnamefont {Malaescu}}, and \bibinfo
  {author} {\bibfnamefont {Z.}~\bibnamefont {Zhang}},\ }\href {\doibase
  10.1140/epjc/s10052-020-7792-2} {\bibfield  {journal} {\bibinfo  {journal}
  {Eur. Phys. J. C}\ }\textbf {\bibinfo {volume} {80}},\ \bibinfo {pages} {241}
  (\bibinfo {year} {2020})},\ \bibinfo {note} {[Erratum: Eur. Phys. J. C {\bf
  80}, 410 (2020)]},\ \Eprint {http://arxiv.org/abs/1908.00921}
  {arXiv:1908.00921 [hep-ph]}\BibitemShut {NoStop}%
\bibitem [{\citenamefont {Keshavarzi}\ \emph
  {et~al.}(2020{\natexlab{a}})\citenamefont {Keshavarzi}, \citenamefont
  {Nomura},\ and\ \citenamefont {Teubner}}]{Keshavarzi:2019abf}%
  \BibitemOpen
  \bibfield  {author} {\bibinfo {author} {\bibfnamefont {A.}~\bibnamefont
  {Keshavarzi}}, \bibinfo {author} {\bibfnamefont {D.}~\bibnamefont {Nomura}},
  and \bibinfo {author} {\bibfnamefont {T.}~\bibnamefont {Teubner}},\ }\href
  {\doibase 10.1103/PhysRevD.101.014029} {\bibfield  {journal} {\bibinfo
  {journal} {Phys. Rev. D}\ }\textbf {\bibinfo {volume} {101}},\ \bibinfo
  {pages} {014029} (\bibinfo {year} {2020}{\natexlab{a}})},\ \Eprint
  {http://arxiv.org/abs/1911.00367} {arXiv:1911.00367 [hep-ph]}\BibitemShut
  {NoStop}%
\bibitem [{\citenamefont {Hoid}\ \emph {et~al.}(2020)\citenamefont {Hoid},
  \citenamefont {Hoferichter},\ and\ \citenamefont {Kubis}}]{Hoid:2020xjs}%
  \BibitemOpen
  \bibfield  {author} {\bibinfo {author} {\bibfnamefont {B.-L.}\ \bibnamefont
  {Hoid}}, \bibinfo {author} {\bibfnamefont {M.}~\bibnamefont {Hoferichter}},
  and \bibinfo {author} {\bibfnamefont {B.}~\bibnamefont {Kubis}},\ }\href
  {\doibase 10.1140/epjc/s10052-020-08550-2} {\bibfield  {journal} {\bibinfo
  {journal} {Eur. Phys. J. C}\ }\textbf {\bibinfo {volume} {80}},\ \bibinfo
  {pages} {988} (\bibinfo {year} {2020})},\ \Eprint
  {http://arxiv.org/abs/2007.12696} {arXiv:2007.12696 [hep-ph]}\BibitemShut
  {NoStop}%
\bibitem [{\citenamefont {Kurz}\ \emph {et~al.}(2014)\citenamefont {Kurz},
  \citenamefont {Liu}, \citenamefont {Marquard},\ and\ \citenamefont
  {Steinhauser}}]{Kurz:2014wya}%
  \BibitemOpen
  \bibfield  {author} {\bibinfo {author} {\bibfnamefont {A.}~\bibnamefont
  {Kurz}}, \bibinfo {author} {\bibfnamefont {T.}~\bibnamefont {Liu}}, \bibinfo
  {author} {\bibfnamefont {P.}~\bibnamefont {Marquard}}, and \bibinfo {author}
  {\bibfnamefont {M.}~\bibnamefont {Steinhauser}},\ }\href {\doibase
  10.1016/j.physletb.2014.05.043} {\bibfield  {journal} {\bibinfo  {journal}
  {Phys. Lett. B}\ }\textbf {\bibinfo {volume} {734}},\ \bibinfo {pages} {144}
  (\bibinfo {year} {2014})},\ \Eprint {http://arxiv.org/abs/1403.6400}
  {arXiv:1403.6400 [hep-ph]}\BibitemShut {NoStop}%
\bibitem [{\citenamefont {Melnikov}\ and\ \citenamefont
  {Vainshtein}(2004)}]{Melnikov:2003xd}%
  \BibitemOpen
  \bibfield  {author} {\bibinfo {author} {\bibfnamefont {K.}~\bibnamefont
  {Melnikov}} and \bibinfo {author} {\bibfnamefont {A.}~\bibnamefont
  {Vainshtein}},\ }\href {\doibase 10.1103/PhysRevD.70.113006} {\bibfield
  {journal} {\bibinfo  {journal} {Phys. Rev. D}\ }\textbf {\bibinfo {volume}
  {70}},\ \bibinfo {pages} {113006} (\bibinfo {year} {2004})},\ \Eprint
  {http://arxiv.org/abs/hep-ph/0312226} {arXiv:hep-ph/0312226}\BibitemShut
  {NoStop}%
\bibitem [{\citenamefont {Colangelo}\ \emph
  {et~al.}(2014{\natexlab{a}})\citenamefont {Colangelo}, \citenamefont
  {Hoferichter}, \citenamefont {Procura},\ and\ \citenamefont
  {Stoffer}}]{Colangelo:2014dfa}%
  \BibitemOpen
  \bibfield  {author} {\bibinfo {author} {\bibfnamefont {G.}~\bibnamefont
  {Colangelo}}, \bibinfo {author} {\bibfnamefont {M.}~\bibnamefont
  {Hoferichter}}, \bibinfo {author} {\bibfnamefont {M.}~\bibnamefont
  {Procura}}, and \bibinfo {author} {\bibfnamefont {P.}~\bibnamefont
  {Stoffer}},\ }\href {\doibase 10.1007/JHEP09(2014)091} {\bibfield  {journal}
  {\bibinfo  {journal} {JHEP}\ }\textbf {\bibinfo {volume} {09}},\ \bibinfo
  {pages} {091} (\bibinfo {year} {2014}{\natexlab{a}})},\ \Eprint
  {http://arxiv.org/abs/1402.7081} {arXiv:1402.7081 [hep-ph]}\BibitemShut
  {NoStop}%
\bibitem [{\citenamefont {Colangelo}\ \emph
  {et~al.}(2014{\natexlab{b}})\citenamefont {Colangelo}, \citenamefont
  {Hoferichter}, \citenamefont {Kubis}, \citenamefont {Procura},\ and\
  \citenamefont {Stoffer}}]{Colangelo:2014pva}%
  \BibitemOpen
  \bibfield  {author} {\bibinfo {author} {\bibfnamefont {G.}~\bibnamefont
  {Colangelo}}, \bibinfo {author} {\bibfnamefont {M.}~\bibnamefont
  {Hoferichter}}, \bibinfo {author} {\bibfnamefont {B.}~\bibnamefont {Kubis}},
  \bibinfo {author} {\bibfnamefont {M.}~\bibnamefont {Procura}}, and \bibinfo
  {author} {\bibfnamefont {P.}~\bibnamefont {Stoffer}},\ }\href {\doibase
  10.1016/j.physletb.2014.09.021} {\bibfield  {journal} {\bibinfo  {journal}
  {Phys. Lett. B}\ }\textbf {\bibinfo {volume} {738}},\ \bibinfo {pages} {6}
  (\bibinfo {year} {2014}{\natexlab{b}})},\ \Eprint
  {http://arxiv.org/abs/1408.2517} {arXiv:1408.2517 [hep-ph]}\BibitemShut
  {NoStop}%
\bibitem [{\citenamefont {Colangelo}\ \emph {et~al.}(2015)\citenamefont
  {Colangelo}, \citenamefont {Hoferichter}, \citenamefont {Procura},\ and\
  \citenamefont {Stoffer}}]{Colangelo:2015ama}%
  \BibitemOpen
  \bibfield  {author} {\bibinfo {author} {\bibfnamefont {G.}~\bibnamefont
  {Colangelo}}, \bibinfo {author} {\bibfnamefont {M.}~\bibnamefont
  {Hoferichter}}, \bibinfo {author} {\bibfnamefont {M.}~\bibnamefont
  {Procura}}, and \bibinfo {author} {\bibfnamefont {P.}~\bibnamefont
  {Stoffer}},\ }\href {\doibase 10.1007/JHEP09(2015)074} {\bibfield  {journal}
  {\bibinfo  {journal} {JHEP}\ }\textbf {\bibinfo {volume} {09}},\ \bibinfo
  {pages} {074} (\bibinfo {year} {2015})},\ \Eprint
  {http://arxiv.org/abs/1506.01386} {arXiv:1506.01386 [hep-ph]}\BibitemShut
  {NoStop}%
\bibitem [{\citenamefont {Masjuan}\ and\ \citenamefont
  {S{\'a}nchez-Puertas}(2017)}]{Masjuan:2017tvw}%
  \BibitemOpen
  \bibfield  {author} {\bibinfo {author} {\bibfnamefont {P.}~\bibnamefont
  {Masjuan}} and \bibinfo {author} {\bibfnamefont {P.}~\bibnamefont
  {S{\'a}nchez-Puertas}},\ }\href {\doibase 10.1103/PhysRevD.95.054026}
  {\bibfield  {journal} {\bibinfo  {journal} {Phys. Rev. D}\ }\textbf {\bibinfo
  {volume} {95}},\ \bibinfo {pages} {054026} (\bibinfo {year} {2017})},\
  \Eprint {http://arxiv.org/abs/1701.05829} {arXiv:1701.05829
  [hep-ph]}\BibitemShut {NoStop}%
\bibitem [{\citenamefont {Colangelo}\ \emph
  {et~al.}(2017{\natexlab{a}})\citenamefont {Colangelo}, \citenamefont
  {Hoferichter}, \citenamefont {Procura},\ and\ \citenamefont
  {Stoffer}}]{Colangelo:2017qdm}%
  \BibitemOpen
  \bibfield  {author} {\bibinfo {author} {\bibfnamefont {G.}~\bibnamefont
  {Colangelo}}, \bibinfo {author} {\bibfnamefont {M.}~\bibnamefont
  {Hoferichter}}, \bibinfo {author} {\bibfnamefont {M.}~\bibnamefont
  {Procura}}, and \bibinfo {author} {\bibfnamefont {P.}~\bibnamefont
  {Stoffer}},\ }\href {\doibase 10.1103/PhysRevLett.118.232001} {\bibfield
  {journal} {\bibinfo  {journal} {Phys. Rev. Lett.}\ }\textbf {\bibinfo
  {volume} {118}},\ \bibinfo {pages} {232001} (\bibinfo {year}
  {2017}{\natexlab{a}})},\ \Eprint {http://arxiv.org/abs/1701.06554}
  {arXiv:1701.06554 [hep-ph]}\BibitemShut {NoStop}%
\bibitem [{\citenamefont {Colangelo}\ \emph
  {et~al.}(2017{\natexlab{b}})\citenamefont {Colangelo}, \citenamefont
  {Hoferichter}, \citenamefont {Procura},\ and\ \citenamefont
  {Stoffer}}]{Colangelo:2017fiz}%
  \BibitemOpen
  \bibfield  {author} {\bibinfo {author} {\bibfnamefont {G.}~\bibnamefont
  {Colangelo}}, \bibinfo {author} {\bibfnamefont {M.}~\bibnamefont
  {Hoferichter}}, \bibinfo {author} {\bibfnamefont {M.}~\bibnamefont
  {Procura}}, and \bibinfo {author} {\bibfnamefont {P.}~\bibnamefont
  {Stoffer}},\ }\href {\doibase 10.1007/JHEP04(2017)161} {\bibfield  {journal}
  {\bibinfo  {journal} {JHEP}\ }\textbf {\bibinfo {volume} {04}},\ \bibinfo
  {pages} {161} (\bibinfo {year} {2017}{\natexlab{b}})},\ \Eprint
  {http://arxiv.org/abs/1702.07347} {arXiv:1702.07347 [hep-ph]}\BibitemShut
  {NoStop}%
\bibitem [{\citenamefont {Hoferichter}\ \emph
  {et~al.}(2018{\natexlab{a}})\citenamefont {Hoferichter}, \citenamefont
  {Hoid}, \citenamefont {Kubis}, \citenamefont {Leupold},\ and\ \citenamefont
  {Schneider}}]{Hoferichter:2018dmo}%
  \BibitemOpen
  \bibfield  {author} {\bibinfo {author} {\bibfnamefont {M.}~\bibnamefont
  {Hoferichter}}, \bibinfo {author} {\bibfnamefont {B.-L.}\ \bibnamefont
  {Hoid}}, \bibinfo {author} {\bibfnamefont {B.}~\bibnamefont {Kubis}},
  \bibinfo {author} {\bibfnamefont {S.}~\bibnamefont {Leupold}}, and \bibinfo
  {author} {\bibfnamefont {S.~P.}\ \bibnamefont {Schneider}},\ }\href {\doibase
  10.1103/PhysRevLett.121.112002} {\bibfield  {journal} {\bibinfo  {journal}
  {Phys. Rev. Lett.}\ }\textbf {\bibinfo {volume} {121}},\ \bibinfo {pages}
  {112002} (\bibinfo {year} {2018}{\natexlab{a}})},\ \Eprint
  {http://arxiv.org/abs/1805.01471} {arXiv:1805.01471 [hep-ph]}\BibitemShut
  {NoStop}%
\bibitem [{\citenamefont {Hoferichter}\ \emph
  {et~al.}(2018{\natexlab{b}})\citenamefont {Hoferichter}, \citenamefont
  {Hoid}, \citenamefont {Kubis}, \citenamefont {Leupold},\ and\ \citenamefont
  {Schneider}}]{Hoferichter:2018kwz}%
  \BibitemOpen
  \bibfield  {author} {\bibinfo {author} {\bibfnamefont {M.}~\bibnamefont
  {Hoferichter}}, \bibinfo {author} {\bibfnamefont {B.-L.}\ \bibnamefont
  {Hoid}}, \bibinfo {author} {\bibfnamefont {B.}~\bibnamefont {Kubis}},
  \bibinfo {author} {\bibfnamefont {S.}~\bibnamefont {Leupold}}, and \bibinfo
  {author} {\bibfnamefont {S.~P.}\ \bibnamefont {Schneider}},\ }\href {\doibase
  10.1007/JHEP10(2018)141} {\bibfield  {journal} {\bibinfo  {journal} {JHEP}\
  }\textbf {\bibinfo {volume} {10}},\ \bibinfo {pages} {141} (\bibinfo {year}
  {2018}{\natexlab{b}})},\ \Eprint {http://arxiv.org/abs/1808.04823}
  {arXiv:1808.04823 [hep-ph]}\BibitemShut {NoStop}%
\bibitem [{\citenamefont {G\'erardin}\ \emph {et~al.}(2019)\citenamefont
  {G\'erardin}, \citenamefont {Meyer},\ and\ \citenamefont
  {Nyffeler}}]{Gerardin:2019vio}%
  \BibitemOpen
  \bibfield  {author} {\bibinfo {author} {\bibfnamefont {A.}~\bibnamefont
  {G\'erardin}}, \bibinfo {author} {\bibfnamefont {H.~B.}\ \bibnamefont
  {Meyer}}, and \bibinfo {author} {\bibfnamefont {A.}~\bibnamefont
  {Nyffeler}},\ }\href {\doibase 10.1103/PhysRevD.100.034520} {\bibfield
  {journal} {\bibinfo  {journal} {Phys. Rev. D}\ }\textbf {\bibinfo {volume}
  {100}},\ \bibinfo {pages} {034520} (\bibinfo {year} {2019})},\ \Eprint
  {http://arxiv.org/abs/1903.09471} {arXiv:1903.09471 [hep-lat]}\BibitemShut
  {NoStop}%
\bibitem [{\citenamefont {Bijnens}\ \emph {et~al.}(2019)\citenamefont
  {Bijnens}, \citenamefont {Hermansson-Truedsson},\ and\ \citenamefont
  {Rodr\'\i{}guez-S\'anchez}}]{Bijnens:2019ghy}%
  \BibitemOpen
  \bibfield  {author} {\bibinfo {author} {\bibfnamefont {J.}~\bibnamefont
  {Bijnens}}, \bibinfo {author} {\bibfnamefont {N.}~\bibnamefont
  {Hermansson-Truedsson}}, and \bibinfo {author} {\bibfnamefont
  {A.}~\bibnamefont {Rodr\'\i{}guez-S\'anchez}},\ }\href {\doibase
  10.1016/j.physletb.2019.134994} {\bibfield  {journal} {\bibinfo  {journal}
  {Phys. Lett. B}\ }\textbf {\bibinfo {volume} {798}},\ \bibinfo {pages}
  {134994} (\bibinfo {year} {2019})},\ \Eprint
  {http://arxiv.org/abs/1908.03331} {arXiv:1908.03331 [hep-ph]}\BibitemShut
  {NoStop}%
\bibitem [{\citenamefont {Colangelo}\ \emph
  {et~al.}(2020{\natexlab{a}})\citenamefont {Colangelo}, \citenamefont
  {Hagelstein}, \citenamefont {Hoferichter}, \citenamefont {Laub},\ and\
  \citenamefont {Stoffer}}]{Colangelo:2019lpu}%
  \BibitemOpen
  \bibfield  {author} {\bibinfo {author} {\bibfnamefont {G.}~\bibnamefont
  {Colangelo}}, \bibinfo {author} {\bibfnamefont {F.}~\bibnamefont
  {Hagelstein}}, \bibinfo {author} {\bibfnamefont {M.}~\bibnamefont
  {Hoferichter}}, \bibinfo {author} {\bibfnamefont {L.}~\bibnamefont {Laub}},
  and \bibinfo {author} {\bibfnamefont {P.}~\bibnamefont {Stoffer}},\ }\href
  {\doibase 10.1103/PhysRevD.101.051501} {\bibfield  {journal} {\bibinfo
  {journal} {Phys. Rev. D}\ }\textbf {\bibinfo {volume} {101}},\ \bibinfo
  {pages} {051501} (\bibinfo {year} {2020}{\natexlab{a}})},\ \Eprint
  {http://arxiv.org/abs/1910.11881} {arXiv:1910.11881 [hep-ph]}\BibitemShut
  {NoStop}%
\bibitem [{\citenamefont {Colangelo}\ \emph
  {et~al.}(2020{\natexlab{b}})\citenamefont {Colangelo}, \citenamefont
  {Hagelstein}, \citenamefont {Hoferichter}, \citenamefont {Laub},\ and\
  \citenamefont {Stoffer}}]{Colangelo:2019uex}%
  \BibitemOpen
  \bibfield  {author} {\bibinfo {author} {\bibfnamefont {G.}~\bibnamefont
  {Colangelo}}, \bibinfo {author} {\bibfnamefont {F.}~\bibnamefont
  {Hagelstein}}, \bibinfo {author} {\bibfnamefont {M.}~\bibnamefont
  {Hoferichter}}, \bibinfo {author} {\bibfnamefont {L.}~\bibnamefont {Laub}},
  and \bibinfo {author} {\bibfnamefont {P.}~\bibnamefont {Stoffer}},\ }\href
  {\doibase 10.1007/JHEP03(2020)101} {\bibfield  {journal} {\bibinfo  {journal}
  {JHEP}\ }\textbf {\bibinfo {volume} {03}},\ \bibinfo {pages} {101} (\bibinfo
  {year} {2020}{\natexlab{b}})},\ \Eprint {http://arxiv.org/abs/1910.13432}
  {arXiv:1910.13432 [hep-ph]}\BibitemShut {NoStop}%
\bibitem [{\citenamefont {Blum}\ \emph {et~al.}(2020)\citenamefont {Blum},
  \citenamefont {Christ}, \citenamefont {Hayakawa}, \citenamefont {Izubuchi},
  \citenamefont {Jin}, \citenamefont {Jung},\ and\ \citenamefont
  {Lehner}}]{Blum:2019ugy}%
  \BibitemOpen
  \bibfield  {author} {\bibinfo {author} {\bibfnamefont {T.}~\bibnamefont
  {Blum}}, \bibinfo {author} {\bibfnamefont {N.}~\bibnamefont {Christ}},
  \bibinfo {author} {\bibfnamefont {M.}~\bibnamefont {Hayakawa}}, \bibinfo
  {author} {\bibfnamefont {T.}~\bibnamefont {Izubuchi}}, \bibinfo {author}
  {\bibfnamefont {L.}~\bibnamefont {Jin}}, \bibinfo {author} {\bibfnamefont
  {C.}~\bibnamefont {Jung}}, and \bibinfo {author} {\bibfnamefont
  {C.}~\bibnamefont {Lehner}} (\bibinfo {collaboration} {RBC, UKQCD}),\ }\href
  {\doibase 10.1103/PhysRevLett.124.132002} {\bibfield  {journal} {\bibinfo
  {journal} {Phys. Rev. Lett.}\ }\textbf {\bibinfo {volume} {124}},\ \bibinfo
  {pages} {132002} (\bibinfo {year} {2020})},\ \Eprint
  {http://arxiv.org/abs/1911.08123} {arXiv:1911.08123 [hep-lat]}\BibitemShut
  {NoStop}%
\bibitem [{\citenamefont {Colangelo}\ \emph
  {et~al.}(2014{\natexlab{c}})\citenamefont {Colangelo}, \citenamefont
  {Hoferichter}, \citenamefont {Nyffeler}, \citenamefont {Passera},\ and\
  \citenamefont {Stoffer}}]{Colangelo:2014qya}%
  \BibitemOpen
  \bibfield  {author} {\bibinfo {author} {\bibfnamefont {G.}~\bibnamefont
  {Colangelo}}, \bibinfo {author} {\bibfnamefont {M.}~\bibnamefont
  {Hoferichter}}, \bibinfo {author} {\bibfnamefont {A.}~\bibnamefont
  {Nyffeler}}, \bibinfo {author} {\bibfnamefont {M.}~\bibnamefont {Passera}},
  and \bibinfo {author} {\bibfnamefont {P.}~\bibnamefont {Stoffer}},\ }\href
  {\doibase 10.1016/j.physletb.2014.06.012} {\bibfield  {journal} {\bibinfo
  {journal} {Phys. Lett. B}\ }\textbf {\bibinfo {volume} {735}},\ \bibinfo
  {pages} {90} (\bibinfo {year} {2014}{\natexlab{c}})},\ \Eprint
  {http://arxiv.org/abs/1403.7512} {arXiv:1403.7512 [hep-ph]}\BibitemShut
  {NoStop}%
\bibitem [{\citenamefont {Aoyama}\ \emph {et~al.}(2020)\citenamefont {Aoyama}
  \emph {et~al.}}]{Aoyama:2020ynm}%
  \BibitemOpen
  \bibfield  {author} {\bibinfo {author} {\bibfnamefont {T.}~\bibnamefont
  {Aoyama}}  \emph {et~al.},\ }\href {\doibase 10.1016/j.physrep.2020.07.006}
  {\bibfield  {journal} {\bibinfo  {journal} {Phys. Rept.}\ }\textbf {\bibinfo
  {volume} {887}},\ \bibinfo {pages} {1} (\bibinfo {year} {2020})},\ \Eprint
  {http://arxiv.org/abs/2006.04822} {arXiv:2006.04822 [hep-ph]}\BibitemShut
  {NoStop}%
\bibitem [{\citenamefont {Bouchiat}\ and\ \citenamefont
  {Michel}(1961)}]{Bouchiat:1961lbg}%
  \BibitemOpen
  \bibfield  {author} {\bibinfo {author} {\bibfnamefont {C.}~\bibnamefont
  {Bouchiat}} and \bibinfo {author} {\bibfnamefont {L.}~\bibnamefont
  {Michel}},\ }\href {\doibase 10.1051/jphysrad:01961002202012101} {\bibfield
  {journal} {\bibinfo  {journal} {J. Phys. Radium}\ }\textbf {\bibinfo {volume}
  {22}},\ \bibinfo {pages} {121} (\bibinfo {year} {1961})}\BibitemShut
  {NoStop}%
\bibitem [{\citenamefont {Brodsky}\ and\ \citenamefont
  {de~Rafael}(1968)}]{Brodsky:1967sr}%
  \BibitemOpen
  \bibfield  {author} {\bibinfo {author} {\bibfnamefont {S.~J.}\ \bibnamefont
  {Brodsky}} and \bibinfo {author} {\bibfnamefont {E.}~\bibnamefont
  {de~Rafael}},\ }\href {\doibase 10.1103/PhysRev.168.1620} {\bibfield
  {journal} {\bibinfo  {journal} {Phys. Rev.}\ }\textbf {\bibinfo {volume}
  {168}},\ \bibinfo {pages} {1620} (\bibinfo {year} {1968})}\BibitemShut
  {NoStop}%
\bibitem [{\citenamefont {Borsanyi}\ \emph {et~al.}(2021)\citenamefont
  {Borsanyi} \emph {et~al.}}]{Borsanyi:2020mff}%
  \BibitemOpen
  \bibfield  {author} {\bibinfo {author} {\bibfnamefont {S.}~\bibnamefont
  {Borsanyi}}  \emph {et~al.} (\bibinfo {collaboration} {BMWc}),\ }\href
  {\doibase 10.1038/s41586-021-03418-1} {\bibfield  {journal} {\bibinfo
  {journal} {Nature}\ }\textbf {\bibinfo {volume} {593}},\ \bibinfo {pages}
  {51} (\bibinfo {year} {2021})},\ \Eprint {http://arxiv.org/abs/2002.12347}
  {arXiv:2002.12347 [hep-lat]}\BibitemShut {NoStop}%
\bibitem [{\citenamefont {Ignatov}\ \emph {et~al.}(2023)\citenamefont {Ignatov}
  \emph {et~al.}}]{CMD-3:2023alj}%
  \BibitemOpen
  \bibfield  {author} {\bibinfo {author} {\bibfnamefont {F.~V.}\ \bibnamefont
  {Ignatov}}  \emph {et~al.} (\bibinfo {collaboration} {CMD-3}),\ }\href@noop
  {} {\  (\bibinfo {year} {2023})},\ \Eprint {http://arxiv.org/abs/2302.08834}
  {arXiv:2302.08834 [hep-ex]}\BibitemShut {NoStop}%
\bibitem [{Muo()}]{MuonInitiative}%
  \BibitemOpen
  \href@noop {} {\enquote {\bibinfo {title} {{Muon $g-2$ Theory Initiative}},}\
  }\bibinfo {howpublished}
  {\url{https://muon-gm2-theory.illinois.edu/}}\BibitemShut {NoStop}%
\bibitem [{\citenamefont {Bardeen}\ and\ \citenamefont
  {Tung}(1968)}]{Bardeen:1969aw}%
  \BibitemOpen
  \bibfield  {author} {\bibinfo {author} {\bibfnamefont {W.~A.}\ \bibnamefont
  {Bardeen}} and \bibinfo {author} {\bibfnamefont {W.~K.}\ \bibnamefont
  {Tung}},\ }\href {\doibase 10.1103/PhysRev.173.1423} {\bibfield  {journal}
  {\bibinfo  {journal} {Phys. Rev.}\ }\textbf {\bibinfo {volume} {173}},\
  \bibinfo {pages} {1423} (\bibinfo {year} {1968})},\ \bibinfo {note}
  {[Erratum: Phys. Rev. D {\bf 4}, 3229 (1971)]}\BibitemShut {NoStop}%
\bibitem [{\citenamefont {Tarrach}(1975)}]{Tarrach:1975tu}%
  \BibitemOpen
  \bibfield  {author} {\bibinfo {author} {\bibfnamefont {R.}~\bibnamefont
  {Tarrach}},\ }\href {\doibase 10.1007/BF02894857} {\bibfield  {journal}
  {\bibinfo  {journal} {Nuovo Cim. A}\ }\textbf {\bibinfo {volume} {28}},\
  \bibinfo {pages} {409} (\bibinfo {year} {1975})}\BibitemShut {NoStop}%
\bibitem [{\citenamefont {Pauk}\ and\ \citenamefont
  {Vanderhaeghen}(2014)}]{Pauk:2014rta}%
  \BibitemOpen
  \bibfield  {author} {\bibinfo {author} {\bibfnamefont {V.}~\bibnamefont
  {Pauk}} and \bibinfo {author} {\bibfnamefont {M.}~\bibnamefont
  {Vanderhaeghen}},\ }\href {\doibase 10.1140/epjc/s10052-014-3008-y}
  {\bibfield  {journal} {\bibinfo  {journal} {Eur. Phys. J. C}\ }\textbf
  {\bibinfo {volume} {74}},\ \bibinfo {pages} {3008} (\bibinfo {year}
  {2014})},\ \Eprint {http://arxiv.org/abs/1401.0832} {arXiv:1401.0832
  [hep-ph]}\BibitemShut {NoStop}%
\bibitem [{\citenamefont {Danilkin}\ and\ \citenamefont
  {Vanderhaeghen}(2017)}]{Danilkin:2016hnh}%
  \BibitemOpen
  \bibfield  {author} {\bibinfo {author} {\bibfnamefont {I.}~\bibnamefont
  {Danilkin}} and \bibinfo {author} {\bibfnamefont {M.}~\bibnamefont
  {Vanderhaeghen}},\ }\href {\doibase 10.1103/PhysRevD.95.014019} {\bibfield
  {journal} {\bibinfo  {journal} {Phys. Rev. D}\ }\textbf {\bibinfo {volume}
  {95}},\ \bibinfo {pages} {014019} (\bibinfo {year} {2017})},\ \Eprint
  {http://arxiv.org/abs/1611.04646} {arXiv:1611.04646 [hep-ph]}\BibitemShut
  {NoStop}%
\bibitem [{\citenamefont {Jegerlehner}(2017)}]{Jegerlehner:2017gek}%
  \BibitemOpen
  \bibfield  {author} {\bibinfo {author} {\bibfnamefont {F.}~\bibnamefont
  {Jegerlehner}},\ }\href {\doibase 10.1007/978-3-319-63577-4} {\emph {\bibinfo
  {title} {{The Anomalous Magnetic Moment of the Muon}}}},\ Vol.\ \bibinfo
  {volume} {274}\ (\bibinfo  {publisher} {Springer},\ \bibinfo {address}
  {Cham},\ \bibinfo {year} {2017})\BibitemShut {NoStop}%
\bibitem [{\citenamefont {Knecht}\ \emph {et~al.}(2018)\citenamefont {Knecht},
  \citenamefont {Narison}, \citenamefont {Rabemananjara},\ and\ \citenamefont
  {Rabetiarivony}}]{Knecht:2018sci}%
  \BibitemOpen
  \bibfield  {author} {\bibinfo {author} {\bibfnamefont {M.}~\bibnamefont
  {Knecht}}, \bibinfo {author} {\bibfnamefont {S.}~\bibnamefont {Narison}},
  \bibinfo {author} {\bibfnamefont {A.}~\bibnamefont {Rabemananjara}}, and
  \bibinfo {author} {\bibfnamefont {D.}~\bibnamefont {Rabetiarivony}},\ }\href
  {\doibase 10.1016/j.physletb.2018.10.048} {\bibfield  {journal} {\bibinfo
  {journal} {Phys. Lett. B}\ }\textbf {\bibinfo {volume} {787}},\ \bibinfo
  {pages} {111} (\bibinfo {year} {2018})},\ \Eprint
  {http://arxiv.org/abs/1808.03848} {arXiv:1808.03848 [hep-ph]}\BibitemShut
  {NoStop}%
\bibitem [{\citenamefont {Eichmann}\ \emph {et~al.}(2020)\citenamefont
  {Eichmann}, \citenamefont {Fischer},\ and\ \citenamefont
  {Williams}}]{Eichmann:2019bqf}%
  \BibitemOpen
  \bibfield  {author} {\bibinfo {author} {\bibfnamefont {G.}~\bibnamefont
  {Eichmann}}, \bibinfo {author} {\bibfnamefont {C.~S.}\ \bibnamefont
  {Fischer}}, and \bibinfo {author} {\bibfnamefont {R.}~\bibnamefont
  {Williams}},\ }\href {\doibase 10.1103/PhysRevD.101.054015} {\bibfield
  {journal} {\bibinfo  {journal} {Phys. Rev. D}\ }\textbf {\bibinfo {volume}
  {101}},\ \bibinfo {pages} {054015} (\bibinfo {year} {2020})},\ \Eprint
  {http://arxiv.org/abs/1910.06795} {arXiv:1910.06795 [hep-ph]}\BibitemShut
  {NoStop}%
\bibitem [{\citenamefont {Roig}\ and\ \citenamefont
  {S{\'a}nchez-Puertas}(2020)}]{Roig:2019reh}%
  \BibitemOpen
  \bibfield  {author} {\bibinfo {author} {\bibfnamefont {P.}~\bibnamefont
  {Roig}} and \bibinfo {author} {\bibfnamefont {P.}~\bibnamefont
  {S{\'a}nchez-Puertas}},\ }\href {\doibase 10.1103/PhysRevD.101.074019}
  {\bibfield  {journal} {\bibinfo  {journal} {Phys. Rev. D}\ }\textbf {\bibinfo
  {volume} {101}},\ \bibinfo {pages} {074019} (\bibinfo {year} {2020})},\
  \Eprint {http://arxiv.org/abs/1910.02881} {arXiv:1910.02881
  [hep-ph]}\BibitemShut {NoStop}%
\bibitem [{\citenamefont {Danilkin}\ \emph {et~al.}(2021)\citenamefont
  {Danilkin}, \citenamefont {Hoferichter},\ and\ \citenamefont
  {Stoffer}}]{Danilkin:2021icn}%
  \BibitemOpen
  \bibfield  {author} {\bibinfo {author} {\bibfnamefont {I.}~\bibnamefont
  {Danilkin}}, \bibinfo {author} {\bibfnamefont {M.}~\bibnamefont
  {Hoferichter}}, and \bibinfo {author} {\bibfnamefont {P.}~\bibnamefont
  {Stoffer}},\ }\href {\doibase 10.1016/j.physletb.2021.136502} {\bibfield
  {journal} {\bibinfo  {journal} {Phys. Lett. B}\ }\textbf {\bibinfo {volume}
  {820}},\ \bibinfo {pages} {136502} (\bibinfo {year} {2021})},\ \Eprint
  {http://arxiv.org/abs/2105.01666} {arXiv:2105.01666 [hep-ph]}\BibitemShut
  {NoStop}%
\bibitem [{\citenamefont {Stamen}\ \emph {et~al.}(2022)\citenamefont {Stamen},
  \citenamefont {Hariharan}, \citenamefont {Hoferichter}, \citenamefont
  {Kubis},\ and\ \citenamefont {Stoffer}}]{Stamen:2022uqh}%
  \BibitemOpen
  \bibfield  {author} {\bibinfo {author} {\bibfnamefont {D.}~\bibnamefont
  {Stamen}}, \bibinfo {author} {\bibfnamefont {D.}~\bibnamefont {Hariharan}},
  \bibinfo {author} {\bibfnamefont {M.}~\bibnamefont {Hoferichter}}, \bibinfo
  {author} {\bibfnamefont {B.}~\bibnamefont {Kubis}}, and \bibinfo {author}
  {\bibfnamefont {P.}~\bibnamefont {Stoffer}},\ }\href {\doibase
  10.1140/epjc/s10052-022-10348-3} {\bibfield  {journal} {\bibinfo  {journal}
  {Eur. Phys. J. C}\ }\textbf {\bibinfo {volume} {82}},\ \bibinfo {pages} {432}
  (\bibinfo {year} {2022})},\ \Eprint {http://arxiv.org/abs/2202.11106}
  {arXiv:2202.11106 [hep-ph]}\BibitemShut {NoStop}%
\bibitem [{\citenamefont {Miramontes}\ \emph {et~al.}(2022)\citenamefont
  {Miramontes}, \citenamefont {Bashir}, \citenamefont {Raya},\ and\
  \citenamefont {Roig}}]{Miramontes:2021exi}%
  \BibitemOpen
  \bibfield  {author} {\bibinfo {author} {\bibfnamefont {A.}~\bibnamefont
  {Miramontes}}, \bibinfo {author} {\bibfnamefont {A.}~\bibnamefont {Bashir}},
  \bibinfo {author} {\bibfnamefont {K.}~\bibnamefont {Raya}}, and \bibinfo
  {author} {\bibfnamefont {P.}~\bibnamefont {Roig}},\ }\href {\doibase
  10.1103/PhysRevD.105.074013} {\bibfield  {journal} {\bibinfo  {journal}
  {Phys. Rev. D}\ }\textbf {\bibinfo {volume} {105}},\ \bibinfo {pages}
  {074013} (\bibinfo {year} {2022})},\ \Eprint
  {http://arxiv.org/abs/2112.13916} {arXiv:2112.13916 [hep-ph]}\BibitemShut
  {NoStop}%
\bibitem [{\citenamefont {Leutgeb}\ and\ \citenamefont
  {Rebhan}(2020)}]{Leutgeb:2019gbz}%
  \BibitemOpen
  \bibfield  {author} {\bibinfo {author} {\bibfnamefont {J.}~\bibnamefont
  {Leutgeb}} and \bibinfo {author} {\bibfnamefont {A.}~\bibnamefont {Rebhan}},\
  }\href {\doibase 10.1103/PhysRevD.101.114015} {\bibfield  {journal} {\bibinfo
   {journal} {Phys. Rev. D}\ }\textbf {\bibinfo {volume} {101}},\ \bibinfo
  {pages} {114015} (\bibinfo {year} {2020})},\ \Eprint
  {http://arxiv.org/abs/1912.01596} {arXiv:1912.01596 [hep-ph]}\BibitemShut
  {NoStop}%
\bibitem [{\citenamefont {Cappiello}\ \emph {et~al.}(2020)\citenamefont
  {Cappiello}, \citenamefont {Cat\`a}, \citenamefont {D'Ambrosio},
  \citenamefont {Greynat},\ and\ \citenamefont {Iyer}}]{Cappiello:2019hwh}%
  \BibitemOpen
  \bibfield  {author} {\bibinfo {author} {\bibfnamefont {L.}~\bibnamefont
  {Cappiello}}, \bibinfo {author} {\bibfnamefont {O.}~\bibnamefont {Cat\`a}},
  \bibinfo {author} {\bibfnamefont {G.}~\bibnamefont {D'Ambrosio}}, \bibinfo
  {author} {\bibfnamefont {D.}~\bibnamefont {Greynat}}, and \bibinfo {author}
  {\bibfnamefont {A.}~\bibnamefont {Iyer}},\ }\href {\doibase
  10.1103/PhysRevD.102.016009} {\bibfield  {journal} {\bibinfo  {journal}
  {Phys. Rev. D}\ }\textbf {\bibinfo {volume} {102}},\ \bibinfo {pages}
  {016009} (\bibinfo {year} {2020})},\ \Eprint
  {http://arxiv.org/abs/1912.02779} {arXiv:1912.02779 [hep-ph]}\BibitemShut
  {NoStop}%
\bibitem [{\citenamefont {Leutgeb}\ \emph {et~al.}(2023)\citenamefont
  {Leutgeb}, \citenamefont {Mager},\ and\ \citenamefont
  {Rebhan}}]{Leutgeb:2022lqw}%
  \BibitemOpen
  \bibfield  {author} {\bibinfo {author} {\bibfnamefont {J.}~\bibnamefont
  {Leutgeb}}, \bibinfo {author} {\bibfnamefont {J.}~\bibnamefont {Mager}}, and
  \bibinfo {author} {\bibfnamefont {A.}~\bibnamefont {Rebhan}},\ }\href
  {\doibase 10.1103/PhysRevD.107.054021} {\bibfield  {journal} {\bibinfo
  {journal} {Phys. Rev. D}\ }\textbf {\bibinfo {volume} {107}},\ \bibinfo
  {pages} {054021} (\bibinfo {year} {2023})},\ \Eprint
  {http://arxiv.org/abs/2211.16562} {arXiv:2211.16562 [hep-ph]}\BibitemShut
  {NoStop}%
\bibitem [{\citenamefont {Colangelo}\ \emph
  {et~al.}(2021{\natexlab{a}})\citenamefont {Colangelo}, \citenamefont
  {Hagelstein}, \citenamefont {Hoferichter}, \citenamefont {Laub},\ and\
  \citenamefont {Stoffer}}]{Colangelo:2021nkr}%
  \BibitemOpen
  \bibfield  {author} {\bibinfo {author} {\bibfnamefont {G.}~\bibnamefont
  {Colangelo}}, \bibinfo {author} {\bibfnamefont {F.}~\bibnamefont
  {Hagelstein}}, \bibinfo {author} {\bibfnamefont {M.}~\bibnamefont
  {Hoferichter}}, \bibinfo {author} {\bibfnamefont {L.}~\bibnamefont {Laub}},
  and \bibinfo {author} {\bibfnamefont {P.}~\bibnamefont {Stoffer}},\ }\href
  {\doibase 10.1140/epjc/s10052-021-09513-x} {\bibfield  {journal} {\bibinfo
  {journal} {Eur. Phys. J. C}\ }\textbf {\bibinfo {volume} {81}},\ \bibinfo
  {pages} {702} (\bibinfo {year} {2021}{\natexlab{a}})},\ \Eprint
  {http://arxiv.org/abs/2106.13222} {arXiv:2106.13222 [hep-ph]}\BibitemShut
  {NoStop}%
\bibitem [{\citenamefont {Hoferichter}\ and\ \citenamefont
  {Stoffer}(2020)}]{Hoferichter:2020lap}%
  \BibitemOpen
  \bibfield  {author} {\bibinfo {author} {\bibfnamefont {M.}~\bibnamefont
  {Hoferichter}} and \bibinfo {author} {\bibfnamefont {P.}~\bibnamefont
  {Stoffer}},\ }\href {\doibase 10.1007/JHEP05(2020)159} {\bibfield  {journal}
  {\bibinfo  {journal} {JHEP}\ }\textbf {\bibinfo {volume} {05}},\ \bibinfo
  {pages} {159} (\bibinfo {year} {2020})},\ \Eprint
  {http://arxiv.org/abs/2004.06127} {arXiv:2004.06127 [hep-ph]}\BibitemShut
  {NoStop}%
\bibitem [{\citenamefont {Zanke}\ \emph {et~al.}(2021)\citenamefont {Zanke},
  \citenamefont {Hoferichter},\ and\ \citenamefont {Kubis}}]{Zanke:2021wiq}%
  \BibitemOpen
  \bibfield  {author} {\bibinfo {author} {\bibfnamefont {M.}~\bibnamefont
  {Zanke}}, \bibinfo {author} {\bibfnamefont {M.}~\bibnamefont {Hoferichter}},
  and \bibinfo {author} {\bibfnamefont {B.}~\bibnamefont {Kubis}},\ }\href
  {\doibase 10.1007/JHEP07(2021)106} {\bibfield  {journal} {\bibinfo  {journal}
  {JHEP}\ }\textbf {\bibinfo {volume} {07}},\ \bibinfo {pages} {106} (\bibinfo
  {year} {2021})},\ \Eprint {http://arxiv.org/abs/2103.09829} {arXiv:2103.09829
  [hep-ph]}\BibitemShut {NoStop}%
\bibitem [{\citenamefont {Hoferichter}\ \emph
  {et~al.}(2023{\natexlab{a}})\citenamefont {Hoferichter}, \citenamefont
  {Kubis},\ and\ \citenamefont {Zanke}}]{Hoferichter:2023tgp}%
  \BibitemOpen
  \bibfield  {author} {\bibinfo {author} {\bibfnamefont {M.}~\bibnamefont
  {Hoferichter}}, \bibinfo {author} {\bibfnamefont {B.}~\bibnamefont {Kubis}},
  and \bibinfo {author} {\bibfnamefont {M.}~\bibnamefont {Zanke}},\ }\href@noop
  {} {\  (\bibinfo {year} {2023}{\natexlab{a}})},\ \Eprint
  {http://arxiv.org/abs/2307.14413} {arXiv:2307.14413 [hep-ph]}\BibitemShut
  {NoStop}%
\bibitem [{\citenamefont {Hoferichter}\ and\ \citenamefont
  {Stoffer}(2019)}]{Hoferichter:2019nlq}%
  \BibitemOpen
  \bibfield  {author} {\bibinfo {author} {\bibfnamefont {M.}~\bibnamefont
  {Hoferichter}} and \bibinfo {author} {\bibfnamefont {P.}~\bibnamefont
  {Stoffer}},\ }\href {\doibase 10.1007/JHEP07(2019)073} {\bibfield  {journal}
  {\bibinfo  {journal} {JHEP}\ }\textbf {\bibinfo {volume} {07}},\ \bibinfo
  {pages} {073} (\bibinfo {year} {2019})},\ \Eprint
  {http://arxiv.org/abs/1905.13198} {arXiv:1905.13198 [hep-ph]}\BibitemShut
  {NoStop}%
\bibitem [{\citenamefont {Danilkin}\ \emph {et~al.}(2020)\citenamefont
  {Danilkin}, \citenamefont {Deineka},\ and\ \citenamefont
  {Vanderhaeghen}}]{Danilkin:2019opj}%
  \BibitemOpen
  \bibfield  {author} {\bibinfo {author} {\bibfnamefont {I.}~\bibnamefont
  {Danilkin}}, \bibinfo {author} {\bibfnamefont {O.}~\bibnamefont {Deineka}},
  and \bibinfo {author} {\bibfnamefont {M.}~\bibnamefont {Vanderhaeghen}},\
  }\href {\doibase 10.1103/PhysRevD.101.054008} {\bibfield  {journal} {\bibinfo
   {journal} {Phys. Rev. D}\ }\textbf {\bibinfo {volume} {101}},\ \bibinfo
  {pages} {054008} (\bibinfo {year} {2020})},\ \Eprint
  {http://arxiv.org/abs/1909.04158} {arXiv:1909.04158 [hep-ph]}\BibitemShut
  {NoStop}%
\bibitem [{\citenamefont {L\"udtke}\ \emph {et~al.}(2023)\citenamefont
  {L\"udtke}, \citenamefont {Procura},\ and\ \citenamefont
  {Stoffer}}]{Ludtke:2023hvz}%
  \BibitemOpen
  \bibfield  {author} {\bibinfo {author} {\bibfnamefont {J.}~\bibnamefont
  {L\"udtke}}, \bibinfo {author} {\bibfnamefont {M.}~\bibnamefont {Procura}},
  and \bibinfo {author} {\bibfnamefont {P.}~\bibnamefont {Stoffer}},\ }\href
  {\doibase 10.1007/JHEP04(2023)125} {\bibfield  {journal} {\bibinfo  {journal}
  {JHEP}\ }\textbf {\bibinfo {volume} {04}},\ \bibinfo {pages} {125} (\bibinfo
  {year} {2023})},\ \Eprint {http://arxiv.org/abs/2302.12264} {arXiv:2302.12264
  [hep-ph]}\BibitemShut {NoStop}%
\bibitem [{\citenamefont {L\"udtke}(2023)}]{Ludtke:2023vgd}%
  \BibitemOpen
  \bibfield  {author} {\bibinfo {author} {\bibfnamefont {J.}~\bibnamefont
  {L\"udtke}},\ }\href {\doibase 10.25365/thesis.73530} {Ph.D. thesis},\
  \bibinfo  {school} {Vienna University} (\bibinfo {year} {2023})\BibitemShut
  {NoStop}%
\bibitem [{\citenamefont {Bijnens}\ \emph {et~al.}(2020)\citenamefont
  {Bijnens}, \citenamefont {Hermansson-Truedsson}, \citenamefont {Laub},\ and\
  \citenamefont {Rodr\'iguez-S\'anchez}}]{Bijnens:2020xnl}%
  \BibitemOpen
  \bibfield  {author} {\bibinfo {author} {\bibfnamefont {J.}~\bibnamefont
  {Bijnens}}, \bibinfo {author} {\bibfnamefont {N.}~\bibnamefont
  {Hermansson-Truedsson}}, \bibinfo {author} {\bibfnamefont {L.}~\bibnamefont
  {Laub}}, and \bibinfo {author} {\bibfnamefont {A.}~\bibnamefont
  {Rodr\'iguez-S\'anchez}},\ }\href {\doibase 10.1007/JHEP10(2020)203}
  {\bibfield  {journal} {\bibinfo  {journal} {JHEP}\ }\textbf {\bibinfo
  {volume} {10}},\ \bibinfo {pages} {203} (\bibinfo {year} {2020})},\ \Eprint
  {http://arxiv.org/abs/2008.13487} {arXiv:2008.13487 [hep-ph]}\BibitemShut
  {NoStop}%
\bibitem [{\citenamefont {Bijnens}\ \emph {et~al.}(2021)\citenamefont
  {Bijnens}, \citenamefont {Hermansson-Truedsson}, \citenamefont {Laub},\ and\
  \citenamefont {Rodr\'iguez-S\'anchez}}]{Bijnens:2021jqo}%
  \BibitemOpen
  \bibfield  {author} {\bibinfo {author} {\bibfnamefont {J.}~\bibnamefont
  {Bijnens}}, \bibinfo {author} {\bibfnamefont {N.}~\bibnamefont
  {Hermansson-Truedsson}}, \bibinfo {author} {\bibfnamefont {L.}~\bibnamefont
  {Laub}}, and \bibinfo {author} {\bibfnamefont {A.}~\bibnamefont
  {Rodr\'iguez-S\'anchez}},\ }\href {\doibase 10.1007/JHEP04(2021)240}
  {\bibfield  {journal} {\bibinfo  {journal} {JHEP}\ }\textbf {\bibinfo
  {volume} {04}},\ \bibinfo {pages} {240} (\bibinfo {year} {2021})},\ \Eprint
  {http://arxiv.org/abs/2101.09169} {arXiv:2101.09169 [hep-ph]}\BibitemShut
  {NoStop}%
\bibitem [{\citenamefont {Bijnens}\ \emph {et~al.}(2023)\citenamefont
  {Bijnens}, \citenamefont {Hermansson-Truedsson},\ and\ \citenamefont
  {Rodr\'\i{}guez-S\'anchez}}]{Bijnens:2022itw}%
  \BibitemOpen
  \bibfield  {author} {\bibinfo {author} {\bibfnamefont {J.}~\bibnamefont
  {Bijnens}}, \bibinfo {author} {\bibfnamefont {N.}~\bibnamefont
  {Hermansson-Truedsson}}, and \bibinfo {author} {\bibfnamefont
  {A.}~\bibnamefont {Rodr\'\i{}guez-S\'anchez}},\ }\href {\doibase
  10.1007/JHEP02(2023)167} {\bibfield  {journal} {\bibinfo  {journal} {JHEP}\
  }\textbf {\bibinfo {volume} {02}},\ \bibinfo {pages} {167} (\bibinfo {year}
  {2023})},\ \Eprint {http://arxiv.org/abs/2211.17183} {arXiv:2211.17183
  [hep-ph]}\BibitemShut {NoStop}%
\bibitem [{\citenamefont {Knecht}(2020)}]{Knecht:2020xyr}%
  \BibitemOpen
  \bibfield  {author} {\bibinfo {author} {\bibfnamefont {M.}~\bibnamefont
  {Knecht}},\ }\href {\doibase 10.1007/JHEP08(2020)056} {\bibfield  {journal}
  {\bibinfo  {journal} {JHEP}\ }\textbf {\bibinfo {volume} {08}},\ \bibinfo
  {pages} {056} (\bibinfo {year} {2020})},\ \Eprint
  {http://arxiv.org/abs/2005.09929} {arXiv:2005.09929 [hep-ph]}\BibitemShut
  {NoStop}%
\bibitem [{\citenamefont {Masjuan}\ \emph {et~al.}(2022)\citenamefont
  {Masjuan}, \citenamefont {Roig},\ and\ \citenamefont
  {S{\'a}nchez-Puertas}}]{Masjuan:2020jsf}%
  \BibitemOpen
  \bibfield  {author} {\bibinfo {author} {\bibfnamefont {P.}~\bibnamefont
  {Masjuan}}, \bibinfo {author} {\bibfnamefont {P.}~\bibnamefont {Roig}}, and
  \bibinfo {author} {\bibfnamefont {P.}~\bibnamefont {S{\'a}nchez-Puertas}},\
  }\href {\doibase 10.1088/1361-6471/ac3892} {\bibfield  {journal} {\bibinfo
  {journal} {J. Phys. G}\ }\textbf {\bibinfo {volume} {49}},\ \bibinfo {pages}
  {015002} (\bibinfo {year} {2022})},\ \Eprint
  {http://arxiv.org/abs/2005.11761} {arXiv:2005.11761 [hep-ph]}\BibitemShut
  {NoStop}%
\bibitem [{\citenamefont {L\"udtke}\ and\ \citenamefont
  {Procura}(2020)}]{Ludtke:2020moa}%
  \BibitemOpen
  \bibfield  {author} {\bibinfo {author} {\bibfnamefont {J.}~\bibnamefont
  {L\"udtke}} and \bibinfo {author} {\bibfnamefont {M.}~\bibnamefont
  {Procura}},\ }\href {\doibase 10.1140/epjc/s10052-020-08611-6} {\bibfield
  {journal} {\bibinfo  {journal} {Eur. Phys. J. C}\ }\textbf {\bibinfo {volume}
  {80}},\ \bibinfo {pages} {1108} (\bibinfo {year} {2020})},\ \Eprint
  {http://arxiv.org/abs/2006.00007} {arXiv:2006.00007 [hep-ph]}\BibitemShut
  {NoStop}%
\bibitem [{\citenamefont {Holz}\ \emph {et~al.}(2021)\citenamefont {Holz},
  \citenamefont {Plenter}, \citenamefont {Xiao}, \citenamefont {Dato},
  \citenamefont {Hanhart}, \citenamefont {Kubis}, \citenamefont {Mei\ss{}ner},\
  and\ \citenamefont {Wirzba}}]{Holz:2015tcg}%
  \BibitemOpen
  \bibfield  {author} {\bibinfo {author} {\bibfnamefont {S.}~\bibnamefont
  {Holz}}, \bibinfo {author} {\bibfnamefont {J.}~\bibnamefont {Plenter}},
  \bibinfo {author} {\bibfnamefont {C.~W.}\ \bibnamefont {Xiao}}, \bibinfo
  {author} {\bibfnamefont {T.}~\bibnamefont {Dato}}, \bibinfo {author}
  {\bibfnamefont {C.}~\bibnamefont {Hanhart}}, \bibinfo {author} {\bibfnamefont
  {B.}~\bibnamefont {Kubis}}, \bibinfo {author} {\bibfnamefont {U.-G.}\
  \bibnamefont {Mei\ss{}ner}}, and \bibinfo {author} {\bibfnamefont
  {A.}~\bibnamefont {Wirzba}},\ }\href {\doibase
  10.1140/epjc/s10052-021-09661-0} {\bibfield  {journal} {\bibinfo  {journal}
  {Eur. Phys. J. C}\ }\textbf {\bibinfo {volume} {81}},\ \bibinfo {pages}
  {1002} (\bibinfo {year} {2021})},\ \Eprint {http://arxiv.org/abs/1509.02194}
  {arXiv:1509.02194 [hep-ph]}\BibitemShut {NoStop}%
\bibitem [{\citenamefont {Holz}\ \emph {et~al.}(2022)\citenamefont {Holz},
  \citenamefont {Hanhart}, \citenamefont {Hoferichter},\ and\ \citenamefont
  {Kubis}}]{Holz:2022hwz}%
  \BibitemOpen
  \bibfield  {author} {\bibinfo {author} {\bibfnamefont {S.}~\bibnamefont
  {Holz}}, \bibinfo {author} {\bibfnamefont {C.}~\bibnamefont {Hanhart}},
  \bibinfo {author} {\bibfnamefont {M.}~\bibnamefont {Hoferichter}}, and
  \bibinfo {author} {\bibfnamefont {B.}~\bibnamefont {Kubis}},\ }\href
  {\doibase 10.1140/epjc/s10052-022-10247-7} {\bibfield  {journal} {\bibinfo
  {journal} {Eur. Phys. J. C}\ }\textbf {\bibinfo {volume} {82}},\ \bibinfo
  {pages} {434} (\bibinfo {year} {2022})},\ \bibinfo {note} {[Addendum: Eur.
  Phys. J. C {\bf 82}, 1159 (2022)]},\ \Eprint
  {http://arxiv.org/abs/2202.05846} {arXiv:2202.05846 [hep-ph]}\BibitemShut
  {NoStop}%
\bibitem [{\citenamefont {Alexandrou}\ \emph {et~al.}(2022)\citenamefont
  {Alexandrou} \emph {et~al.}}]{Alexandrou:2022qyf}%
  \BibitemOpen
  \bibfield  {author} {\bibinfo {author} {\bibfnamefont {C.}~\bibnamefont
  {Alexandrou}}  \emph {et~al.} (\bibinfo {collaboration} {ETM}),\ }\href@noop
  {} {\  (\bibinfo {year} {2022})},\ \Eprint {http://arxiv.org/abs/2212.06704}
  {arXiv:2212.06704 [hep-lat]}\BibitemShut {NoStop}%
\bibitem [{\citenamefont {G\'erardin}\ \emph {et~al.}(2023)\citenamefont
  {G\'erardin} \emph {et~al.}}]{Gerardin:2023naa}%
  \BibitemOpen
  \bibfield  {author} {\bibinfo {author} {\bibfnamefont {A.}~\bibnamefont
  {G\'erardin}}  \emph {et~al.} (\bibinfo {collaboration} {BMWc}),\ }\href@noop
  {} {\  (\bibinfo {year} {2023})},\ \Eprint {http://arxiv.org/abs/2305.04570}
  {arXiv:2305.04570 [hep-lat]}\BibitemShut {NoStop}%
\bibitem [{\citenamefont {Chao}\ \emph {et~al.}(2021)\citenamefont {Chao},
  \citenamefont {Hudspith}, \citenamefont {G\'erardin}, \citenamefont {Green},
  \citenamefont {Meyer},\ and\ \citenamefont {Ottnad}}]{Chao:2021tvp}%
  \BibitemOpen
  \bibfield  {author} {\bibinfo {author} {\bibfnamefont {E.-H.}\ \bibnamefont
  {Chao}}, \bibinfo {author} {\bibfnamefont {R.~J.}\ \bibnamefont {Hudspith}},
  \bibinfo {author} {\bibfnamefont {A.}~\bibnamefont {G\'erardin}}, \bibinfo
  {author} {\bibfnamefont {J.~R.}\ \bibnamefont {Green}}, \bibinfo {author}
  {\bibfnamefont {H.~B.}\ \bibnamefont {Meyer}}, and \bibinfo {author}
  {\bibfnamefont {K.}~\bibnamefont {Ottnad}},\ }\href {\doibase
  10.1140/epjc/s10052-021-09455-4} {\bibfield  {journal} {\bibinfo  {journal}
  {Eur. Phys. J. C}\ }\textbf {\bibinfo {volume} {81}},\ \bibinfo {pages} {651}
  (\bibinfo {year} {2021})},\ \Eprint {http://arxiv.org/abs/2104.02632}
  {arXiv:2104.02632 [hep-lat]}\BibitemShut {NoStop}%
\bibitem [{\citenamefont {Chao}\ \emph {et~al.}(2022)\citenamefont {Chao},
  \citenamefont {Hudspith}, \citenamefont {G\'erardin}, \citenamefont {Green},\
  and\ \citenamefont {Meyer}}]{Chao:2022xzg}%
  \BibitemOpen
  \bibfield  {author} {\bibinfo {author} {\bibfnamefont {E.-H.}\ \bibnamefont
  {Chao}}, \bibinfo {author} {\bibfnamefont {R.~J.}\ \bibnamefont {Hudspith}},
  \bibinfo {author} {\bibfnamefont {A.}~\bibnamefont {G\'erardin}}, \bibinfo
  {author} {\bibfnamefont {J.~R.}\ \bibnamefont {Green}}, and \bibinfo {author}
  {\bibfnamefont {H.~B.}\ \bibnamefont {Meyer}},\ }\href {\doibase
  10.1140/epjc/s10052-022-10589-2} {\bibfield  {journal} {\bibinfo  {journal}
  {Eur. Phys. J. C}\ }\textbf {\bibinfo {volume} {82}},\ \bibinfo {pages} {664}
  (\bibinfo {year} {2022})},\ \Eprint {http://arxiv.org/abs/2204.08844}
  {arXiv:2204.08844 [hep-lat]}\BibitemShut {NoStop}%
\bibitem [{\citenamefont {Blum}\ \emph
  {et~al.}(2023{\natexlab{a}})\citenamefont {Blum}, \citenamefont {Christ},
  \citenamefont {Hayakawa}, \citenamefont {Izubuchi}, \citenamefont {Jin},
  \citenamefont {Jung}, \citenamefont {Lehner},\ and\ \citenamefont
  {Tu}}]{Blum:2023vlm}%
  \BibitemOpen
  \bibfield  {author} {\bibinfo {author} {\bibfnamefont {T.}~\bibnamefont
  {Blum}}, \bibinfo {author} {\bibfnamefont {N.}~\bibnamefont {Christ}},
  \bibinfo {author} {\bibfnamefont {M.}~\bibnamefont {Hayakawa}}, \bibinfo
  {author} {\bibfnamefont {T.}~\bibnamefont {Izubuchi}}, \bibinfo {author}
  {\bibfnamefont {L.}~\bibnamefont {Jin}}, \bibinfo {author} {\bibfnamefont
  {C.}~\bibnamefont {Jung}}, \bibinfo {author} {\bibfnamefont {C.}~\bibnamefont
  {Lehner}}, and \bibinfo {author} {\bibfnamefont {C.}~\bibnamefont {Tu}}
  (\bibinfo {collaboration} {RBC, UKQCD}),\ }\href@noop {} {\  (\bibinfo {year}
  {2023}{\natexlab{a}})},\ \Eprint {http://arxiv.org/abs/2304.04423}
  {arXiv:2304.04423 [hep-lat]}\BibitemShut {NoStop}%
\bibitem [{\citenamefont {Colangelo}\ \emph
  {et~al.}(2022{\natexlab{a}})\citenamefont {Colangelo} \emph
  {et~al.}}]{Colangelo:2022jxc}%
  \BibitemOpen
  \bibfield  {author} {\bibinfo {author} {\bibfnamefont {G.}~\bibnamefont
  {Colangelo}}  \emph {et~al.},\ }\href@noop {} {\  (\bibinfo {year}
  {2022}{\natexlab{a}})},\ \Eprint {http://arxiv.org/abs/2203.15810}
  {arXiv:2203.15810 [hep-ph]}\BibitemShut {NoStop}%
\bibitem [{\citenamefont {Blum}\ \emph {et~al.}(2018)\citenamefont {Blum} \emph
  {et~al.}}]{RBC:2018dos}%
  \BibitemOpen
  \bibfield  {author} {\bibinfo {author} {\bibfnamefont {T.}~\bibnamefont
  {Blum}}  \emph {et~al.} (\bibinfo {collaboration} {RBC, UKQCD}),\ }\href
  {\doibase 10.1103/PhysRevLett.121.022003} {\bibfield  {journal} {\bibinfo
  {journal} {Phys. Rev. Lett.}\ }\textbf {\bibinfo {volume} {121}},\ \bibinfo
  {pages} {022003} (\bibinfo {year} {2018})},\ \Eprint
  {http://arxiv.org/abs/1801.07224} {arXiv:1801.07224 [hep-lat]}\BibitemShut
  {NoStop}%
\bibitem [{\citenamefont {Colangelo}\ \emph
  {et~al.}(2022{\natexlab{b}})\citenamefont {Colangelo}, \citenamefont
  {El-Khadra}, \citenamefont {Hoferichter}, \citenamefont {Keshavarzi},
  \citenamefont {Lehner}, \citenamefont {Stoffer},\ and\ \citenamefont
  {Teubner}}]{Colangelo:2022vok}%
  \BibitemOpen
  \bibfield  {author} {\bibinfo {author} {\bibfnamefont {G.}~\bibnamefont
  {Colangelo}}, \bibinfo {author} {\bibfnamefont {A.~X.}\ \bibnamefont
  {El-Khadra}}, \bibinfo {author} {\bibfnamefont {M.}~\bibnamefont
  {Hoferichter}}, \bibinfo {author} {\bibfnamefont {A.}~\bibnamefont
  {Keshavarzi}}, \bibinfo {author} {\bibfnamefont {C.}~\bibnamefont {Lehner}},
  \bibinfo {author} {\bibfnamefont {P.}~\bibnamefont {Stoffer}}, and \bibinfo
  {author} {\bibfnamefont {T.}~\bibnamefont {Teubner}},\ }\href {\doibase
  10.1016/j.physletb.2022.137313} {\bibfield  {journal} {\bibinfo  {journal}
  {Phys. Lett. B}\ }\textbf {\bibinfo {volume} {833}},\ \bibinfo {pages}
  {137313} (\bibinfo {year} {2022}{\natexlab{b}})},\ \Eprint
  {http://arxiv.org/abs/2205.12963} {arXiv:2205.12963 [hep-ph]}\BibitemShut
  {NoStop}%
\bibitem [{\citenamefont {C\`e}\ \emph
  {et~al.}(2022{\natexlab{a}})\citenamefont {C\`e} \emph
  {et~al.}}]{Ce:2022kxy}%
  \BibitemOpen
  \bibfield  {author} {\bibinfo {author} {\bibfnamefont {M.}~\bibnamefont
  {C\`e}}  \emph {et~al.},\ }\href {\doibase 10.1103/PhysRevD.106.114502}
  {\bibfield  {journal} {\bibinfo  {journal} {Phys. Rev. D}\ }\textbf {\bibinfo
  {volume} {106}},\ \bibinfo {pages} {114502} (\bibinfo {year}
  {2022}{\natexlab{a}})},\ \Eprint {http://arxiv.org/abs/2206.06582}
  {arXiv:2206.06582 [hep-lat]}\BibitemShut {NoStop}%
\bibitem [{\citenamefont {Alexandrou}\ \emph {et~al.}(2023)\citenamefont
  {Alexandrou} \emph {et~al.}}]{ExtendedTwistedMass:2022jpw}%
  \BibitemOpen
  \bibfield  {author} {\bibinfo {author} {\bibfnamefont {C.}~\bibnamefont
  {Alexandrou}}  \emph {et~al.} (\bibinfo {collaboration} {ETM}),\ }\href
  {\doibase 10.1103/PhysRevD.107.074506} {\bibfield  {journal} {\bibinfo
  {journal} {Phys. Rev. D}\ }\textbf {\bibinfo {volume} {107}},\ \bibinfo
  {pages} {074506} (\bibinfo {year} {2023})},\ \Eprint
  {http://arxiv.org/abs/2206.15084} {arXiv:2206.15084 [hep-lat]}\BibitemShut
  {NoStop}%
\bibitem [{\citenamefont {Bazavov}\ \emph {et~al.}(2023)\citenamefont {Bazavov}
  \emph {et~al.}}]{FermilabLatticeHPQCD:2023jof}%
  \BibitemOpen
  \bibfield  {author} {\bibinfo {author} {\bibfnamefont {A.}~\bibnamefont
  {Bazavov}}  \emph {et~al.} (\bibinfo {collaboration} {Fermilab Lattice,
  HPQCD, MILC}),\ }\href {\doibase 10.1103/PhysRevD.107.114514} {\bibfield
  {journal} {\bibinfo  {journal} {Phys. Rev. D}\ }\textbf {\bibinfo {volume}
  {107}},\ \bibinfo {pages} {114514} (\bibinfo {year} {2023})},\ \Eprint
  {http://arxiv.org/abs/2301.08274} {arXiv:2301.08274 [hep-lat]}\BibitemShut
  {NoStop}%
\bibitem [{\citenamefont {Blum}\ \emph
  {et~al.}(2023{\natexlab{b}})\citenamefont {Blum} \emph
  {et~al.}}]{Blum:2023qou}%
  \BibitemOpen
  \bibfield  {author} {\bibinfo {author} {\bibfnamefont {T.}~\bibnamefont
  {Blum}}  \emph {et~al.} (\bibinfo {collaboration} {RBC, UKQCD}),\ }\href@noop
  {} {\  (\bibinfo {year} {2023}{\natexlab{b}})},\ \Eprint
  {http://arxiv.org/abs/2301.08696} {arXiv:2301.08696 [hep-lat]}\BibitemShut
  {NoStop}%
\bibitem [{\citenamefont {Hoferichter}\ \emph
  {et~al.}(2023{\natexlab{b}})\citenamefont {Hoferichter}, \citenamefont
  {Colangelo}, \citenamefont {Hoid}, \citenamefont {Kubis}, \citenamefont
  {de~Elvira}, \citenamefont {Schuh}, \citenamefont {Stamen},\ and\
  \citenamefont {Stoffer}}]{Hoferichter:2023sli}%
  \BibitemOpen
  \bibfield  {author} {\bibinfo {author} {\bibfnamefont {M.}~\bibnamefont
  {Hoferichter}}, \bibinfo {author} {\bibfnamefont {G.}~\bibnamefont
  {Colangelo}}, \bibinfo {author} {\bibfnamefont {B.-L.}\ \bibnamefont {Hoid}},
  \bibinfo {author} {\bibfnamefont {B.}~\bibnamefont {Kubis}}, \bibinfo
  {author} {\bibfnamefont {J.~R.}\ \bibnamefont {de~Elvira}}, \bibinfo {author}
  {\bibfnamefont {D.}~\bibnamefont {Schuh}}, \bibinfo {author} {\bibfnamefont
  {D.}~\bibnamefont {Stamen}}, and \bibinfo {author} {\bibfnamefont
  {P.}~\bibnamefont {Stoffer}},\ }\href@noop {} {\  (\bibinfo {year}
  {2023}{\natexlab{b}})},\ \Eprint {http://arxiv.org/abs/2307.02532}
  {arXiv:2307.02532 [hep-ph]}\BibitemShut {NoStop}%
\bibitem [{\citenamefont {Hoferichter}\ \emph
  {et~al.}(2023{\natexlab{c}})\citenamefont {Hoferichter}, \citenamefont
  {Hoid}, \citenamefont {Kubis},\ and\ \citenamefont
  {Schuh}}]{Hoferichter:2023bjm}%
  \BibitemOpen
  \bibfield  {author} {\bibinfo {author} {\bibfnamefont {M.}~\bibnamefont
  {Hoferichter}}, \bibinfo {author} {\bibfnamefont {B.-L.}\ \bibnamefont
  {Hoid}}, \bibinfo {author} {\bibfnamefont {B.}~\bibnamefont {Kubis}}, and
  \bibinfo {author} {\bibfnamefont {D.}~\bibnamefont {Schuh}},\ }\href@noop {}
  {\  (\bibinfo {year} {2023}{\natexlab{c}})},\ \Eprint
  {http://arxiv.org/abs/2307.02546} {arXiv:2307.02546 [hep-ph]}\BibitemShut
  {NoStop}%
\bibitem [{\citenamefont {Colangelo}\ \emph
  {et~al.}(2022{\natexlab{c}})\citenamefont {Colangelo}, \citenamefont
  {Hoferichter}, \citenamefont {Kubis},\ and\ \citenamefont
  {Stoffer}}]{Colangelo:2022prz}%
  \BibitemOpen
  \bibfield  {author} {\bibinfo {author} {\bibfnamefont {G.}~\bibnamefont
  {Colangelo}}, \bibinfo {author} {\bibfnamefont {M.}~\bibnamefont
  {Hoferichter}}, \bibinfo {author} {\bibfnamefont {B.}~\bibnamefont {Kubis}},
  and \bibinfo {author} {\bibfnamefont {P.}~\bibnamefont {Stoffer}},\ }\href
  {\doibase 10.1007/JHEP10(2022)032} {\bibfield  {journal} {\bibinfo  {journal}
  {JHEP}\ }\textbf {\bibinfo {volume} {10}},\ \bibinfo {pages} {032} (\bibinfo
  {year} {2022}{\natexlab{c}})},\ \Eprint {http://arxiv.org/abs/2208.08993}
  {arXiv:2208.08993 [hep-ph]}\BibitemShut {NoStop}%
\bibitem [{\citenamefont {Colangelo}\ \emph
  {et~al.}(2022{\natexlab{d}})\citenamefont {Colangelo}, \citenamefont
  {Hoferichter}, \citenamefont {Kubis}, \citenamefont {Niehus},\ and\
  \citenamefont {Ruiz~de Elvira}}]{Colangelo:2021moe}%
  \BibitemOpen
  \bibfield  {author} {\bibinfo {author} {\bibfnamefont {G.}~\bibnamefont
  {Colangelo}}, \bibinfo {author} {\bibfnamefont {M.}~\bibnamefont
  {Hoferichter}}, \bibinfo {author} {\bibfnamefont {B.}~\bibnamefont {Kubis}},
  \bibinfo {author} {\bibfnamefont {M.}~\bibnamefont {Niehus}}, and \bibinfo
  {author} {\bibfnamefont {J.}~\bibnamefont {Ruiz~de Elvira}},\ }\href
  {\doibase 10.1016/j.physletb.2021.136852} {\bibfield  {journal} {\bibinfo
  {journal} {Phys. Lett. B}\ }\textbf {\bibinfo {volume} {825}},\ \bibinfo
  {pages} {136852} (\bibinfo {year} {2022}{\natexlab{d}})},\ \Eprint
  {http://arxiv.org/abs/2110.05493} {arXiv:2110.05493 [hep-ph]}\BibitemShut
  {NoStop}%
\bibitem [{\citenamefont {Hoferichter}\ \emph {et~al.}(2022)\citenamefont
  {Hoferichter}, \citenamefont {Colangelo}, \citenamefont {Hoid}, \citenamefont
  {Kubis}, \citenamefont {Ruiz~de Elvira}, \citenamefont {Stamen},\ and\
  \citenamefont {Stoffer}}]{Hoferichter:2022iqe}%
  \BibitemOpen
  \bibfield  {author} {\bibinfo {author} {\bibfnamefont {M.}~\bibnamefont
  {Hoferichter}}, \bibinfo {author} {\bibfnamefont {G.}~\bibnamefont
  {Colangelo}}, \bibinfo {author} {\bibfnamefont {B.-L.}\ \bibnamefont {Hoid}},
  \bibinfo {author} {\bibfnamefont {B.}~\bibnamefont {Kubis}}, \bibinfo
  {author} {\bibfnamefont {J.}~\bibnamefont {Ruiz~de Elvira}}, \bibinfo
  {author} {\bibfnamefont {D.}~\bibnamefont {Stamen}}, and \bibinfo {author}
  {\bibfnamefont {P.}~\bibnamefont {Stoffer}},\ }\href {\doibase
  10.22323/1.430.0316} {\bibfield  {journal} {\bibinfo  {journal} {PoS}\
  }\textbf {\bibinfo {volume} {LATTICE2022}},\ \bibinfo {pages} {316} (\bibinfo
  {year} {2022})},\ \Eprint {http://arxiv.org/abs/2210.11904} {arXiv:2210.11904
  [hep-ph]}\BibitemShut {NoStop}%
\bibitem [{\citenamefont {Campanario}\ \emph {et~al.}(2019)\citenamefont
  {Campanario}, \citenamefont {Czy\.z}, \citenamefont {Gluza}, \citenamefont
  {Jeli\'nski}, \citenamefont {Rodrigo}, \citenamefont {Tracz},\ and\
  \citenamefont {Zhuridov}}]{Campanario:2019mjh}%
  \BibitemOpen
  \bibfield  {author} {\bibinfo {author} {\bibfnamefont {F.}~\bibnamefont
  {Campanario}}, \bibinfo {author} {\bibfnamefont {H.}~\bibnamefont {Czy\.z}},
  \bibinfo {author} {\bibfnamefont {J.}~\bibnamefont {Gluza}}, \bibinfo
  {author} {\bibfnamefont {T.}~\bibnamefont {Jeli\'nski}}, \bibinfo {author}
  {\bibfnamefont {G.}~\bibnamefont {Rodrigo}}, \bibinfo {author} {\bibfnamefont
  {S.}~\bibnamefont {Tracz}}, and \bibinfo {author} {\bibfnamefont
  {D.}~\bibnamefont {Zhuridov}},\ }\href {\doibase 10.1103/PhysRevD.100.076004}
  {\bibfield  {journal} {\bibinfo  {journal} {Phys. Rev. D}\ }\textbf {\bibinfo
  {volume} {100}},\ \bibinfo {pages} {076004} (\bibinfo {year} {2019})},\
  \Eprint {http://arxiv.org/abs/1903.10197} {arXiv:1903.10197
  [hep-ph]}\BibitemShut {NoStop}%
\bibitem [{\citenamefont {Colangelo}\ \emph
  {et~al.}(2022{\natexlab{e}})\citenamefont {Colangelo}, \citenamefont
  {Hoferichter}, \citenamefont {Monnard},\ and\ \citenamefont {Ruiz~de
  Elvira}}]{Colangelo:2022lzg}%
  \BibitemOpen
  \bibfield  {author} {\bibinfo {author} {\bibfnamefont {G.}~\bibnamefont
  {Colangelo}}, \bibinfo {author} {\bibfnamefont {M.}~\bibnamefont
  {Hoferichter}}, \bibinfo {author} {\bibfnamefont {J.}~\bibnamefont
  {Monnard}}, and \bibinfo {author} {\bibfnamefont {J.}~\bibnamefont {Ruiz~de
  Elvira}},\ }\href {\doibase 10.1007/JHEP08(2022)295} {\bibfield  {journal}
  {\bibinfo  {journal} {JHEP}\ }\textbf {\bibinfo {volume} {08}},\ \bibinfo
  {pages} {295} (\bibinfo {year} {2022}{\natexlab{e}})},\ \Eprint
  {http://arxiv.org/abs/2207.03495} {arXiv:2207.03495 [hep-ph]}\BibitemShut
  {NoStop}%
\bibitem [{\citenamefont {Ignatov}\ and\ \citenamefont
  {Lee}(2022)}]{Ignatov:2022iou}%
  \BibitemOpen
  \bibfield  {author} {\bibinfo {author} {\bibfnamefont {F.}~\bibnamefont
  {Ignatov}} and \bibinfo {author} {\bibfnamefont {R.~N.}\ \bibnamefont
  {Lee}},\ }\href {\doibase 10.1016/j.physletb.2022.137283} {\bibfield
  {journal} {\bibinfo  {journal} {Phys. Lett. B}\ }\textbf {\bibinfo {volume}
  {833}},\ \bibinfo {pages} {137283} (\bibinfo {year} {2022})},\ \Eprint
  {http://arxiv.org/abs/2204.12235} {arXiv:2204.12235 [hep-ph]}\BibitemShut
  {NoStop}%
\bibitem [{\citenamefont {Monnard}(2020)}]{JMPhDThesis}%
  \BibitemOpen
  \bibfield  {author} {\bibinfo {author} {\bibfnamefont {J.}~\bibnamefont
  {Monnard}},\ }\href {\doibase https://boristheses.unibe.ch/2825/} {Ph.D.
  thesis},\ \bibinfo  {school} {University of Bern} (\bibinfo {year}
  {2020})\BibitemShut {NoStop}%
\bibitem [{\citenamefont {Abbiendi}\ \emph {et~al.}(2022)\citenamefont
  {Abbiendi} \emph {et~al.}}]{Abbiendi:2022liz}%
  \BibitemOpen
  \bibfield  {author} {\bibinfo {author} {\bibfnamefont {G.}~\bibnamefont
  {Abbiendi}}  \emph {et~al.},\ }\href@noop {} {\  (\bibinfo {year} {2022})},\
  \Eprint {http://arxiv.org/abs/2201.12102} {arXiv:2201.12102
  [hep-ph]}\BibitemShut {NoStop}%
\bibitem [{\citenamefont {Passera}\ \emph {et~al.}(2008)\citenamefont
  {Passera}, \citenamefont {Marciano},\ and\ \citenamefont
  {Sirlin}}]{Passera:2008jk}%
  \BibitemOpen
  \bibfield  {author} {\bibinfo {author} {\bibfnamefont {M.}~\bibnamefont
  {Passera}}, \bibinfo {author} {\bibfnamefont {W.~J.}\ \bibnamefont
  {Marciano}}, and \bibinfo {author} {\bibfnamefont {A.}~\bibnamefont
  {Sirlin}},\ }\href {\doibase 10.1103/PhysRevD.78.013009} {\bibfield
  {journal} {\bibinfo  {journal} {Phys. Rev. D}\ }\textbf {\bibinfo {volume}
  {78}},\ \bibinfo {pages} {013009} (\bibinfo {year} {2008})},\ \Eprint
  {http://arxiv.org/abs/0804.1142} {arXiv:0804.1142 [hep-ph]}\BibitemShut
  {NoStop}%
\bibitem [{\citenamefont {Crivellin}\ \emph {et~al.}(2020)\citenamefont
  {Crivellin}, \citenamefont {Hoferichter}, \citenamefont {Manzari},\ and\
  \citenamefont {Montull}}]{Crivellin:2020zul}%
  \BibitemOpen
  \bibfield  {author} {\bibinfo {author} {\bibfnamefont {A.}~\bibnamefont
  {Crivellin}}, \bibinfo {author} {\bibfnamefont {M.}~\bibnamefont
  {Hoferichter}}, \bibinfo {author} {\bibfnamefont {C.~A.}\ \bibnamefont
  {Manzari}}, and \bibinfo {author} {\bibfnamefont {M.}~\bibnamefont
  {Montull}},\ }\href {\doibase 10.1103/PhysRevLett.125.091801} {\bibfield
  {journal} {\bibinfo  {journal} {Phys. Rev. Lett.}\ }\textbf {\bibinfo
  {volume} {125}},\ \bibinfo {pages} {091801} (\bibinfo {year} {2020})},\
  \Eprint {http://arxiv.org/abs/2003.04886} {arXiv:2003.04886
  [hep-ph]}\BibitemShut {NoStop}%
\bibitem [{\citenamefont {Keshavarzi}\ \emph
  {et~al.}(2020{\natexlab{b}})\citenamefont {Keshavarzi}, \citenamefont
  {Marciano}, \citenamefont {Passera},\ and\ \citenamefont
  {Sirlin}}]{Keshavarzi:2020bfy}%
  \BibitemOpen
  \bibfield  {author} {\bibinfo {author} {\bibfnamefont {A.}~\bibnamefont
  {Keshavarzi}}, \bibinfo {author} {\bibfnamefont {W.~J.}\ \bibnamefont
  {Marciano}}, \bibinfo {author} {\bibfnamefont {M.}~\bibnamefont {Passera}},
  and \bibinfo {author} {\bibfnamefont {A.}~\bibnamefont {Sirlin}},\ }\href
  {\doibase 10.1103/PhysRevD.102.033002} {\bibfield  {journal} {\bibinfo
  {journal} {Phys. Rev. D}\ }\textbf {\bibinfo {volume} {102}},\ \bibinfo
  {pages} {033002} (\bibinfo {year} {2020}{\natexlab{b}})},\ \Eprint
  {http://arxiv.org/abs/2006.12666} {arXiv:2006.12666 [hep-ph]}\BibitemShut
  {NoStop}%
\bibitem [{\citenamefont {Malaescu}\ and\ \citenamefont
  {Schott}(2021)}]{Malaescu:2020zuc}%
  \BibitemOpen
  \bibfield  {author} {\bibinfo {author} {\bibfnamefont {B.}~\bibnamefont
  {Malaescu}} and \bibinfo {author} {\bibfnamefont {M.}~\bibnamefont
  {Schott}},\ }\href {\doibase 10.1140/epjc/s10052-021-08848-9} {\bibfield
  {journal} {\bibinfo  {journal} {Eur. Phys. J. C}\ }\textbf {\bibinfo {volume}
  {81}},\ \bibinfo {pages} {46} (\bibinfo {year} {2021})},\ \Eprint
  {http://arxiv.org/abs/2008.08107} {arXiv:2008.08107 [hep-ph]}\BibitemShut
  {NoStop}%
\bibitem [{\citenamefont {Colangelo}\ \emph
  {et~al.}(2021{\natexlab{b}})\citenamefont {Colangelo}, \citenamefont
  {Hoferichter},\ and\ \citenamefont {Stoffer}}]{Colangelo:2020lcg}%
  \BibitemOpen
  \bibfield  {author} {\bibinfo {author} {\bibfnamefont {G.}~\bibnamefont
  {Colangelo}}, \bibinfo {author} {\bibfnamefont {M.}~\bibnamefont
  {Hoferichter}}, and \bibinfo {author} {\bibfnamefont {P.}~\bibnamefont
  {Stoffer}},\ }\href {\doibase 10.1016/j.physletb.2021.136073} {\bibfield
  {journal} {\bibinfo  {journal} {Phys. Lett. B}\ }\textbf {\bibinfo {volume}
  {814}},\ \bibinfo {pages} {136073} (\bibinfo {year} {2021}{\natexlab{b}})},\
  \Eprint {http://arxiv.org/abs/2010.07943} {arXiv:2010.07943
  [hep-ph]}\BibitemShut {NoStop}%
\bibitem [{\citenamefont {C\`e}\ \emph
  {et~al.}(2022{\natexlab{b}})\citenamefont {C\`e} \emph
  {et~al.}}]{Ce:2022eix}%
  \BibitemOpen
  \bibfield  {author} {\bibinfo {author} {\bibfnamefont {M.}~\bibnamefont
  {C\`e}}  \emph {et~al.},\ }\href {\doibase 10.1007/JHEP08(2022)220}
  {\bibfield  {journal} {\bibinfo  {journal} {JHEP}\ }\textbf {\bibinfo
  {volume} {08}},\ \bibinfo {pages} {220} (\bibinfo {year}
  {2022}{\natexlab{b}})},\ \Eprint {http://arxiv.org/abs/2203.08676}
  {arXiv:2203.08676 [hep-lat]}\BibitemShut {NoStop}%
\bibitem [{\citenamefont {Ananthanarayan}\ \emph {et~al.}(2001)\citenamefont
  {Ananthanarayan}, \citenamefont {Colangelo}, \citenamefont {Gasser},\ and\
  \citenamefont {Leutwyler}}]{Ananthanarayan:2000ht}%
  \BibitemOpen
  \bibfield  {author} {\bibinfo {author} {\bibfnamefont {B.}~\bibnamefont
  {Ananthanarayan}}, \bibinfo {author} {\bibfnamefont {G.}~\bibnamefont
  {Colangelo}}, \bibinfo {author} {\bibfnamefont {J.}~\bibnamefont {Gasser}},
  and \bibinfo {author} {\bibfnamefont {H.}~\bibnamefont {Leutwyler}},\ }\href
  {\doibase 10.1016/S0370-1573(01)00009-6} {\bibfield  {journal} {\bibinfo
  {journal} {Phys. Rept.}\ }\textbf {\bibinfo {volume} {353}},\ \bibinfo
  {pages} {207} (\bibinfo {year} {2001})},\ \Eprint
  {http://arxiv.org/abs/hep-ph/0005297} {arXiv:hep-ph/0005297}\BibitemShut
  {NoStop}%
\bibitem [{\citenamefont {Caprini}\ \emph {et~al.}(2012)\citenamefont
  {Caprini}, \citenamefont {Colangelo},\ and\ \citenamefont
  {Leutwyler}}]{Caprini:2011ky}%
  \BibitemOpen
  \bibfield  {author} {\bibinfo {author} {\bibfnamefont {I.}~\bibnamefont
  {Caprini}}, \bibinfo {author} {\bibfnamefont {G.}~\bibnamefont {Colangelo}},
  and \bibinfo {author} {\bibfnamefont {H.}~\bibnamefont {Leutwyler}},\ }\href
  {\doibase 10.1140/epjc/s10052-012-1860-1} {\bibfield  {journal} {\bibinfo
  {journal} {Eur. Phys. J. C}\ }\textbf {\bibinfo {volume} {72}},\ \bibinfo
  {pages} {1860} (\bibinfo {year} {2012})},\ \Eprint
  {http://arxiv.org/abs/1111.7160} {arXiv:1111.7160 [hep-ph]}\BibitemShut
  {NoStop}%
\bibitem [{\citenamefont {Lees}\ \emph {et~al.}(2012)\citenamefont {Lees} \emph
  {et~al.}}]{Lees:2012cj}%
  \BibitemOpen
  \bibfield  {author} {\bibinfo {author} {\bibfnamefont {J.~P.}\ \bibnamefont
  {Lees}}  \emph {et~al.} (\bibinfo {collaboration} {BaBar}),\ }\href {\doibase
  10.1103/PhysRevD.86.032013} {\bibfield  {journal} {\bibinfo  {journal} {Phys.
  Rev. D}\ }\textbf {\bibinfo {volume} {86}},\ \bibinfo {pages} {032013}
  (\bibinfo {year} {2012})},\ \Eprint {http://arxiv.org/abs/1205.2228}
  {arXiv:1205.2228 [hep-ex]}\BibitemShut {NoStop}%
\bibitem [{\citenamefont {Anastasi}\ \emph {et~al.}(2018)\citenamefont
  {Anastasi} \emph {et~al.}}]{Anastasi:2017eio}%
  \BibitemOpen
  \bibfield  {author} {\bibinfo {author} {\bibfnamefont {A.}~\bibnamefont
  {Anastasi}}  \emph {et~al.} (\bibinfo {collaboration} {KLOE-2}),\ }\href
  {\doibase 10.1007/JHEP03(2018)173} {\bibfield  {journal} {\bibinfo  {journal}
  {JHEP}\ }\textbf {\bibinfo {volume} {03}},\ \bibinfo {pages} {173} (\bibinfo
  {year} {2018})},\ \Eprint {http://arxiv.org/abs/1711.03085} {arXiv:1711.03085
  [hep-ex]}\BibitemShut {NoStop}%
\bibitem [{\citenamefont {Feng}\ \emph {et~al.}(2020)\citenamefont {Feng},
  \citenamefont {Fu},\ and\ \citenamefont {Jin}}]{Feng:2019geu}%
  \BibitemOpen
  \bibfield  {author} {\bibinfo {author} {\bibfnamefont {X.}~\bibnamefont
  {Feng}}, \bibinfo {author} {\bibfnamefont {Y.}~\bibnamefont {Fu}}, and
  \bibinfo {author} {\bibfnamefont {L.-C.}\ \bibnamefont {Jin}},\ }\href
  {\doibase 10.1103/PhysRevD.101.051502} {\bibfield  {journal} {\bibinfo
  {journal} {Phys. Rev. D}\ }\textbf {\bibinfo {volume} {101}},\ \bibinfo
  {pages} {051502} (\bibinfo {year} {2020})},\ \Eprint
  {http://arxiv.org/abs/1911.04064} {arXiv:1911.04064 [hep-lat]}\BibitemShut
  {NoStop}%
\bibitem [{\citenamefont {Wang}\ \emph {et~al.}(2021)\citenamefont {Wang},
  \citenamefont {Liang}, \citenamefont {Draper}, \citenamefont {Liu},\ and\
  \citenamefont {Yang}}]{Wang:2020nbf}%
  \BibitemOpen
  \bibfield  {author} {\bibinfo {author} {\bibfnamefont {G.}~\bibnamefont
  {Wang}}, \bibinfo {author} {\bibfnamefont {J.}~\bibnamefont {Liang}},
  \bibinfo {author} {\bibfnamefont {T.}~\bibnamefont {Draper}}, \bibinfo
  {author} {\bibfnamefont {K.-F.}\ \bibnamefont {Liu}}, and \bibinfo {author}
  {\bibfnamefont {Y.-B.}\ \bibnamefont {Yang}} (\bibinfo {collaboration}
  {$\chi$QCD}),\ }\href {\doibase 10.1103/PhysRevD.104.074502} {\bibfield
  {journal} {\bibinfo  {journal} {Phys. Rev. D}\ }\textbf {\bibinfo {volume}
  {104}},\ \bibinfo {pages} {074502} (\bibinfo {year} {2021})},\ \Eprint
  {http://arxiv.org/abs/2006.05431} {arXiv:2006.05431 [hep-ph]}\BibitemShut
  {NoStop}%
\bibitem [{\citenamefont {Achasov}\ \emph {et~al.}(2021)\citenamefont {Achasov}
  \emph {et~al.}}]{SND:2020nwa}%
  \BibitemOpen
  \bibfield  {author} {\bibinfo {author} {\bibfnamefont {M.~N.}\ \bibnamefont
  {Achasov}}  \emph {et~al.} (\bibinfo {collaboration} {SND}),\ }\href
  {\doibase 10.1007/JHEP01(2021)113} {\bibfield  {journal} {\bibinfo  {journal}
  {JHEP}\ }\textbf {\bibinfo {volume} {01}},\ \bibinfo {pages} {113} (\bibinfo
  {year} {2021})},\ \Eprint {http://arxiv.org/abs/2004.00263} {arXiv:2004.00263
  [hep-ex]}\BibitemShut {NoStop}%
\bibitem [{\citenamefont {Ablikim}\ \emph {et~al.}(2016)\citenamefont {Ablikim}
  \emph {et~al.}}]{Ablikim:2015orh}%
  \BibitemOpen
  \bibfield  {author} {\bibinfo {author} {\bibfnamefont {M.}~\bibnamefont
  {Ablikim}}  \emph {et~al.} (\bibinfo {collaboration} {BESIII}),\ }\href
  {\doibase 10.1016/j.physletb.2015.11.043} {\bibfield  {journal} {\bibinfo
  {journal} {Phys. Lett. B}\ }\textbf {\bibinfo {volume} {753}},\ \bibinfo
  {pages} {629} (\bibinfo {year} {2016})},\ \bibinfo {note} {[Erratum: Phys.
  Lett. B {\bf 812}, 135982 (2021)]},\ \Eprint
  {http://arxiv.org/abs/1507.08188} {arXiv:1507.08188 [hep-ex]}\BibitemShut
  {NoStop}%
\bibitem [{\citenamefont {Achasov}\ \emph {et~al.}(2006)\citenamefont {Achasov}
  \emph {et~al.}}]{Achasov:2006vp}%
  \BibitemOpen
  \bibfield  {author} {\bibinfo {author} {\bibfnamefont {M.~N.}\ \bibnamefont
  {Achasov}}  \emph {et~al.} (\bibinfo {collaboration} {SND}),\ }\href
  {\doibase 10.1134/S106377610609007X} {\bibfield  {journal} {\bibinfo
  {journal} {J. Exp. Theor. Phys.}\ }\textbf {\bibinfo {volume} {103}},\
  \bibinfo {pages} {380} (\bibinfo {year} {2006})},\ \Eprint
  {http://arxiv.org/abs/hep-ex/0605013} {arXiv:hep-ex/0605013}\BibitemShut
  {NoStop}%
\bibitem [{\citenamefont {Akhmetshin}\ \emph {et~al.}(2007)\citenamefont
  {Akhmetshin} \emph {et~al.}}]{Akhmetshin:2006bx}%
  \BibitemOpen
  \bibfield  {author} {\bibinfo {author} {\bibfnamefont {R.~R.}\ \bibnamefont
  {Akhmetshin}}  \emph {et~al.} (\bibinfo {collaboration} {CMD-2}),\ }\href
  {\doibase 10.1016/j.physletb.2007.01.073} {\bibfield  {journal} {\bibinfo
  {journal} {Phys. Lett. B}\ }\textbf {\bibinfo {volume} {648}},\ \bibinfo
  {pages} {28} (\bibinfo {year} {2007})},\ \Eprint
  {http://arxiv.org/abs/hep-ex/0610021} {arXiv:hep-ex/0610021}\BibitemShut
  {NoStop}%
\bibitem [{\citenamefont {Amendolia}\ \emph {et~al.}(1986)\citenamefont
  {Amendolia} \emph {et~al.}}]{Amendolia:1986wj}%
  \BibitemOpen
  \bibfield  {author} {\bibinfo {author} {\bibfnamefont {S.~R.}\ \bibnamefont
  {Amendolia}}  \emph {et~al.} (\bibinfo {collaboration} {NA7}),\ }\href
  {\doibase 10.1016/0550-3213(86)90437-2} {\bibfield  {journal} {\bibinfo
  {journal} {Nucl. Phys. B}\ }\textbf {\bibinfo {volume} {277}},\ \bibinfo
  {pages} {168} (\bibinfo {year} {1986})}\BibitemShut {NoStop}%
\end{thebibliography}%

\end{document}